\documentclass{aa}  

\def \eg {e.g.}

\def \ie {i.e.}
\def \cf {cf.}
\def \lcdm {{\hbox{$\Lambda$CDM}}}
\def \omegam {{\hbox{$\Omega_{\rm m}$}}}
\def \omegal {{\hbox{$\Omega_\Lambda$}}}
\def \hzero {{\hbox{$H_0$}}}
\def \arcmin {\hbox{$^\prime$}}
\def \arcsec {\hbox{$^{\prime\prime}$}}
\def \deg {\hbox{$^\circ$}}
\def \nh {\hbox{$N_{\rm H}$}}

\def \mach {{\hbox{$\mathcal{M}$}}}

\def \compr {{\hbox{$\mathcal{C}$}}}

\def \msun {\hbox{${\rm M_\odot}$}}

\def \mfive {\hbox{$M_{500}$}}

\newcommand{\ergscmsq }{\mbox{erg s$^{-1}$ cm$^{-2}$}}
\newcommand{\ergs }{\mbox{erg s$^{-1}$}}
\newcommand{\kms }{\mbox{km s$^{-1}$}}
\newcommand{\kmsmpc }{\mbox{km s$^{-1}$ Mpc$^{-1}$}}

\newcommand{\kevcmsq }{\mbox{keV cm$^{2}$}}
\newcommand{\jy }{\mbox{Jy}}
\newcommand{\mjyb }{\mbox{mJy beam$^{-1}$}}
\newcommand{\mujyb }{\mbox{$\mu$Jy beam$^{-1}$}}

\newcommand{\whz }{\mbox{W Hz$^{-1}$}}
\newcommand{\acisi }{ACIS-I}
\newcommand{\aciss }{ACIS-S}

\newcommand{\obsid }{ObsID}

\newcommand{\vfaint }{VFAINT}
\newcommand{\uv }{\textit{uv}}
\newcommand{\aips }{\textsc{aips}}
\newcommand{\aipsE }{Astronomical Image Processing System}

\newcommand{\prefactor }{\textsc{prefactor}}

\newcommand{\wsclean }{\textsc{WSClean}}

\newcommand{\killms }{\textsc{killMS}}
\newcommand{\ddfacet }{\textsc{DDFacet}}
\newcommand{\mcmc }{MCMC}
\newcommand{\mcmcE }{Markov chain Monte Carlo}
\newcommand{\xspec }{\textsc{xspec}}
\newcommand{\contbin }{\textsc{contbin}}
\newcommand{\ciao }{\textsc{ciao}}

\newcommand{\caldb }{\textsc{caldb}}

\newcommand{\proffit }{\textsc{proffit}}

\newcommand{\xmm }{{\em XMM-Newton}}
\newcommand{\chandra }{{\em Chandra}}

\newcommand{\athena }{{\em Athena}}

\newcommand{\gmrt }{GMRT}
\newcommand{\gmrtE }{Giant Metrewave Radio Telescope}

\newcommand{\vla }{VLA}
\newcommand{\vlaE }{Very Large Array}

\newcommand{\lofar }{LOFAR}
\newcommand{\lofarE }{LOw Frequency ARray}

\newcommand{\lotss }{LoTSS}
\newcommand{\lotssE }{LOFAR Two-meter Sky Survey}

\newcommand{\panstarrs }{Pan-STARRS}
\newcommand{\panstarrsE }{Panoramic Survey Telescope and Rapid Response System}

\usepackage{graphicx}
\usepackage{amssymb}
\usepackage{multirow,bigdelim}
\usepackage{txfonts}
\usepackage[export]{adjustbox}
\begin{document} 

\title{Nonthermal phenomena in the center of Abell 1775:}
\subtitle{An 800 kpc head-tail, revived fossil plasma and slingshot radio halo}

\authorrunning{A. Botteon et al.} 
\titlerunning{Nonthermal phenomena in the center of Abell 1775}

\author{A. Botteon\inst{\ref{leiden}}, S. Giacintucci\inst{\ref{naval}}, F. Gastaldello\inst{\ref{iasf}}, T. Venturi\inst{\ref{ira}}, G. Brunetti\inst{\ref{ira}}, R. J. van Weeren\inst{\ref{leiden}}, T. W. Shimwell\inst{\ref{astron}, \ref{leiden}}, M. Rossetti\inst{\ref{iasf}}, H. Akamatsu\inst{\ref{sron}}, M. Br\"{u}ggen\inst{\ref{hamburg}}, R. Cassano\inst{\ref{ira}}, V. Cuciti\inst{\ref{hamburg}}, F. de Gasperin\inst{\ref{hamburg}}, A. Drabent\inst{\ref{tls}}, M. Hoeft\inst{\ref{tls}}, S. Mandal\inst{\ref{leiden}}, H. J. A. R\"{o}ttgering\inst{\ref{leiden}}, and C. Tasse\inst{\ref{gepi}, \ref{ratt}}}

\institute{
Leiden Observatory, Leiden University, PO Box 9513, NL-2300 RA Leiden, The Netherlands \label{leiden} \\
\email{botteon@strw.leidenuniv.nl} 
\and
Naval Research Laboratory, 4555 Overlook Avenue SW, Code 7213, Washington, DC 20375, USA \label{naval}
\and
INAF - IASF Milano, via A.~Corti 12, I-20133 Milano, Italy \label{iasf}
\and
INAF - IRA, via P.~Gobetti 101, I-40129 Bologna, Italy \label{ira}
\and
ASTRON, the Netherlands Institute for Radio Astronomy, Postbus 2, NL-7990 AA Dwingeloo, The Netherlands \label{astron}
\and
SRON Netherlands Institute for Space Research, Sorbonnelaan 2, NL-3584 CA Utrecht, The Netherlands \label{sron}
\and
Hamburger Sternwarte, Universit\"{a}t Hamburg, Gojenbergsweg 112, D-21029 Hamburg, Germany \label{hamburg}
\and
Th\"{u}ringer Landessternwarte, Sternwarte 5, D-07778 Tautenburg, Germany \label{tls}
\and
GEPI, Observatoire de Paris, CNRS, Universit\'e Paris Diderot, 5 place Jules Janssen, 92190 Meudon, France \label{gepi}
\and
Centre for Radio Astronomy Techniques and Technologies, Department of Physics and Electronics, Rhodes University, Grahamstown 6140, South Africa \label{ratt}
}

\date{Received XXX; accepted YYY}

\abstract
{Thermal gas in the center of galaxy clusters can show substantial motions that generate surface-brightness and temperature discontinuities known as cold fronts. The motions may be triggered by minor or off-axis mergers that preserve the cool core of the system. The dynamics of the thermal gas can also generate radio emission from the intra-cluster medium (ICM) and impact the evolution of clusters' radio sources.}
{We aim to study the central region of Abell 1775, a system in an ambiguous dynamical state at $z=0.072$ which is known to host an extended head-tail radio galaxy, with the goal of investigating the connection between thermal and nonthermal components in its center.}
{We made use of a deep (100 ks) \chandra\ observation accompanied by \lofar\ 144~MHz, \gmrt\ 235~MHz and 610~MHz, and \vla\ 1.4~GHz radio data.}
{We find a spiral-like pattern in the X-ray surface brightness that is mirrored in the temperature and pseudo-entropy maps. Additionally, we characterize an arc-shaped cold front in the ICM. We interpret these features in the context of a slingshot gas tail scenario. The structure of the head-tail radio galaxy ``breaks'' at the position of the cold front, showing an extension that is detected only at low frequencies, likely due to its steep and curved spectrum. We speculate that particle reacceleration is occurring in the outer region of this tail, which in total covers a projected size of $\sim800$~kpc. We also report the discovery of revived fossil plasma with ultra-steep spectrum radio emission in the cluster core together with a central diffuse radio source that is bounded by the arc-shaped cold front.}
{The results reported in this work demonstrate the interplay between thermal and nonthermal components in the cluster center and the presence of ongoing particle reacceleration in the ICM on different scales.}


\maketitle

\section{Introduction}

The intra-cluster medium (ICM) is a hot and rarified gas that permeates the entire cluster volume and has a temperature on the order of $10^7-10^8$ K and densities of about $10^{-3}-10^{-4}$ particles cm$^{-3}$. In this regime, and for typical ICM metallicities ($0.1-1$ solar), the gas radiates X-rays predominantly through thermal bremsstrahlung emission \citep[\eg,][for a review]{sarazin86rev}. In the hierarchical process of large-scale structure formation, mergers between clusters leave observable imprints in the ICM in the form of shocks, cold fronts, and hydrodynamic instabilities \citep[\eg,][for a review]{markevitch07rev}. In minor and off-axis mergers, right after the pericenter passage, the gas tail of the infalling subcluster overshoots the dark matter peak due to the
rapid decline of ram pressure and produces an arc-shaped slingshot tail \citep[\eg,][]{hallman04, poole06, sheardown19}. The slingshot gas effect has been invoked in the context of gas sloshing in some of the more dynamically relaxed clusters, where the perturbations in the potential well of the more massive system induce a spiral-like motion of its cool core, which, without being disrupted, generates sloshing cold fronts in the ICM \citep{ascasibar06}. Several observations have provided evidence of these processes in the central regions of cool-core clusters \citep[\eg,][]{churazov03, clarke04a2029, lagana10, johnson10, johnson12, blanton11, roediger12a496, sanders16centaurus, chen17, ueda17, douglass18}, and the sloshing scenario is supported by a number of numerical simulations \citep[\eg,][]{tittely05, ascasibar06, zuhone10, roediger11}. \\
\indent
The presence of turbulence in the cores of relaxed clusters is believed to play an important role in the formation of the diffuse synchrotron sources termed radio mini-halos \citep[\eg,][]{gitti02}. Their steep spectrum ($\alpha > 1$, with $S_\nu \propto \nu^{-\alpha}$ where $\alpha$ is the spectral index) emission occupies the cluster cooling region ($r \sim 50-300$ kpc) and is not directly caused by the active galactic nucleus (AGN) that sits in the central cluster galaxy \citep[see][for a recent overview]{giacintucci17, giacintucci19}. Possible evidence for the connection between mini-halos and sloshing motions is provided by observations showing the confinement of some mini-halos inside the region delineated by sloshing cold fronts \citep[\eg,][]{mazzotta08, hlavaceklarrondo13, giacintucci14rxj1720, giacintucci14minihalos, gendronmarsolais17}. Indeed, gas sloshing has been suggested to amplify the magnetic field within the core region \citep[\eg,][]{keshet10distribution, zuhone11} and induce significant turbulent motions in the cluster center \citep{fujita04turbulence, zuhone13}, leading to the reacceleration of the relativistic electrons in mini-halos. Alternatively, it has been proposed that the radio-emitting electrons are generated by inelastic collisions between cosmic-ray protons and thermal protons \citep[\eg,][]{pfrommer04hadronic, fujita07, zandanel14scalings}; in this case, the confinement of the radio emission at cold fronts is expected to be less prominent \citep{zuhone15}. \\
\indent
The dynamics of the ICM can also impact the morphology and spectral properties of tailed cluster radio galaxies. Relativistic electrons in the tails can gain energy via a number of processes, such as turbulent reacceleration due to the growth of Kelvin-Helmholtz instabilities generated by the interaction between the radio plasma and the surrounding medium \citep[\eg,][]{loken95}, adiabatic compression \citep[\eg,][]{ensslin01}, shock reacceleration \citep[\eg,][]{vanweeren17a3411}, or even by more subtle mechanisms that barely balance electrons radiative losses powered by plasma instabilities within the radio tails \citep[\eg,][]{degasperin17gentle}. Among these processes, adiabatic compression of pockets of old nonthermal components due to bulk motions in the ICM is believed to be responsible for the formation of radio emission with very steep spectra ($\alpha>2$) called revived fossil plasma sources or radio phoenixes \citep[\eg,][]{slee01, cohen11, vanweeren11vlss, kale12relics, mandal20}. \\
\indent
The plethora of phenomena described above makes cluster centers unique laboratories to study the physics of diffuse, hot, weakly magnetized cosmic plasmas and their interplay with nonthermal components on a broad range of scales. The analysis of the central region of a cluster is indeed the focus of our study. \\
\indent
Abell 1775 is a galaxy cluster with a richness class of 2 belonging to the Bo\"otes supercluster \citep{einasto97, chowmartinez14}. Optical studies on the velocity distribution of cluster galaxies have revealed that it is a binary \citep{oegerle95, kopylov09} or possibly ternary \citep{zhang11} system. Here, we focus on its main component, Abell 1775B (hereafter simply A1775), located at $z=0.072$, which is known to host two giant elliptical galaxies at its center with a radial-velocity difference of $\sim 1800$ \kms\ \citep{chincarini71, jenner74, hintzen79, kirshner83, zabludoff90}. Both galaxies are active in the radio band, with the southeast one, B1339+266B, whose radial-velocity deviates from the bulk of the velocities of the other cluster members, hosting an extended head-tail radio source that has been studied at different frequencies and resolutions \citep[][]{owen97, giovannini00, giacintucci07, ternidegregory17}. The dynamical state of the cluster is unclear: \citet{lopes18} classified A1775 as a relaxed system based on optical and X-ray information, \citet{andersson09} and \citet{lagana19maps} suggested the presence of dynamical motions in its center from spatially resolved X-ray spectral analysis, and the cool-core metrics used by \citet{andradesantos17} placed it in an ambiguous position between cool-core and non-cool-core systems. A1775 is part of the ROSAT Brightest Cluster Sample with an X-ray luminosity of $1.6 \times 10^{44}$ \ergs\ and a flux of $1.25 \times 10^{-11}$ \ergscmsq\ in the $0.1-2.4$ keV band \citep{ebeling98}. In the second catalog of Planck Sunyaev-Zel'dovich sources \citep{planck16xxvii}, it is reported with the name PSZ2G031.93+78.71 and mass $\mfive = (2.72\pm0.24) \times 10^{14}$ \msun. \\
\indent
In this paper, we report on results based on deep \chandra\ observations and multifrequency radio observations performed with the \lofarE\ (\lofar), \gmrtE\ (\gmrt), and \vlaE\ (\vla) spanning a frequency range from 144~MHz to 1.4~GHz with the aim of studying the central region of A1775 and the impact of the ICM dynamics on the diffuse radio sources in the cluster. We adopted a \lcdm\ cosmology with $\omegal = 0.7$, $\omegam = 0.3,$ and $\hzero = 70$ \kmsmpc, in which 1 arcsec corresponds to 1.372 kpc at $z=0.072$. Unless otherwise stated, reported uncertainties correspond to a 68\% confidence level.

\section{Data reduction}

\subsection{Chandra}\label{sec:chandra}

\begin{figure*}
 \centering
 \includegraphics[width=.75\hsize,trim={0cm 0cm 0cm 0cm},clip]{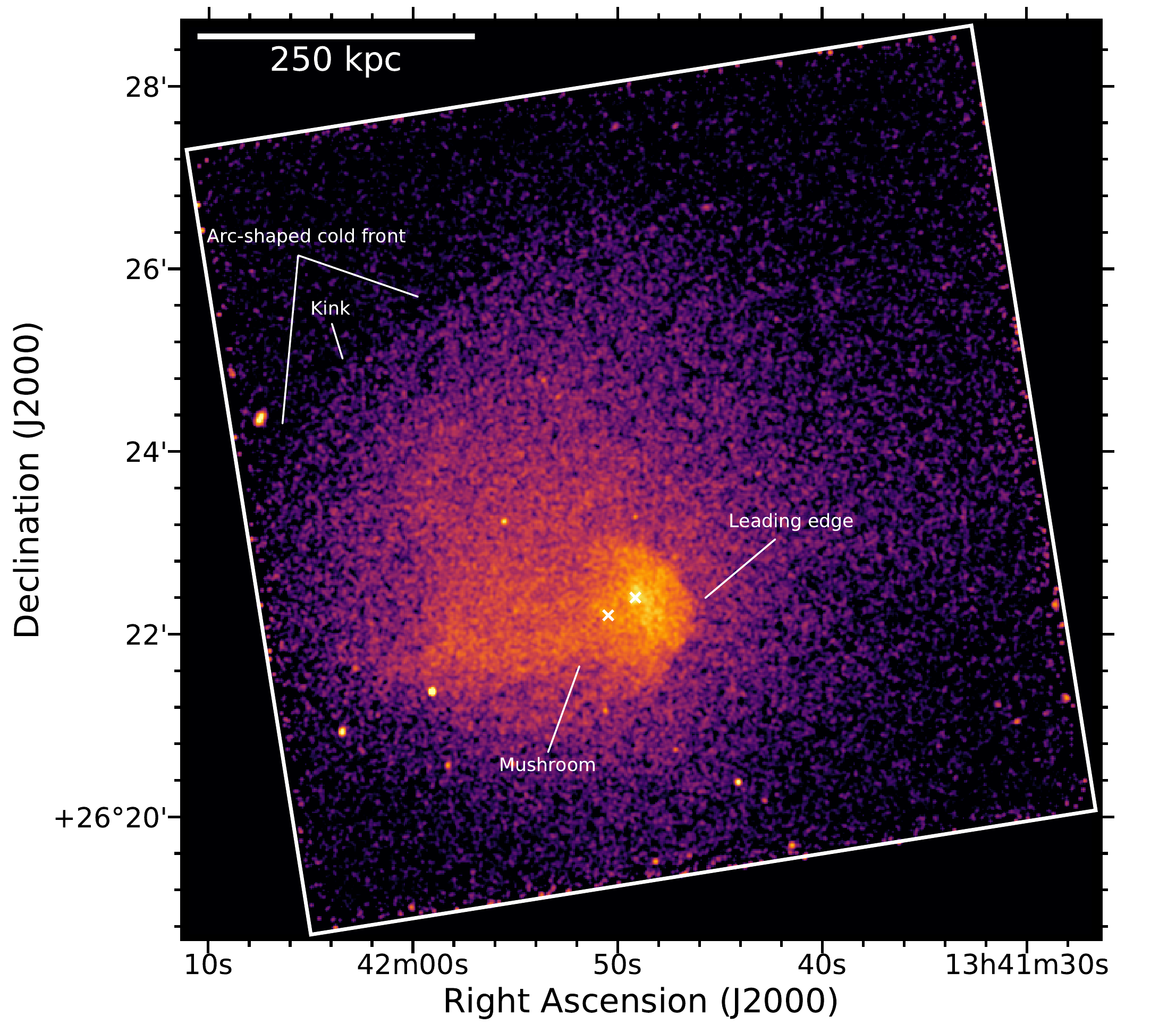}
  \caption{\chandra\ X-ray image of A1775 in the $0.5-2.0$ keV band. The squared box region (for this and subsequent images) denotes the FoV of the \aciss 3 chip. The main features discussed in the text are labeled in the figure. Crosses mark the position of the two giant elliptical galaxies.}
 \label{fig:chandra_image}
\end{figure*}

\begin{figure*}
 \centering
 \includegraphics[width=.5\hsize,trim={0cm 0cm 0cm 0cm},clip]{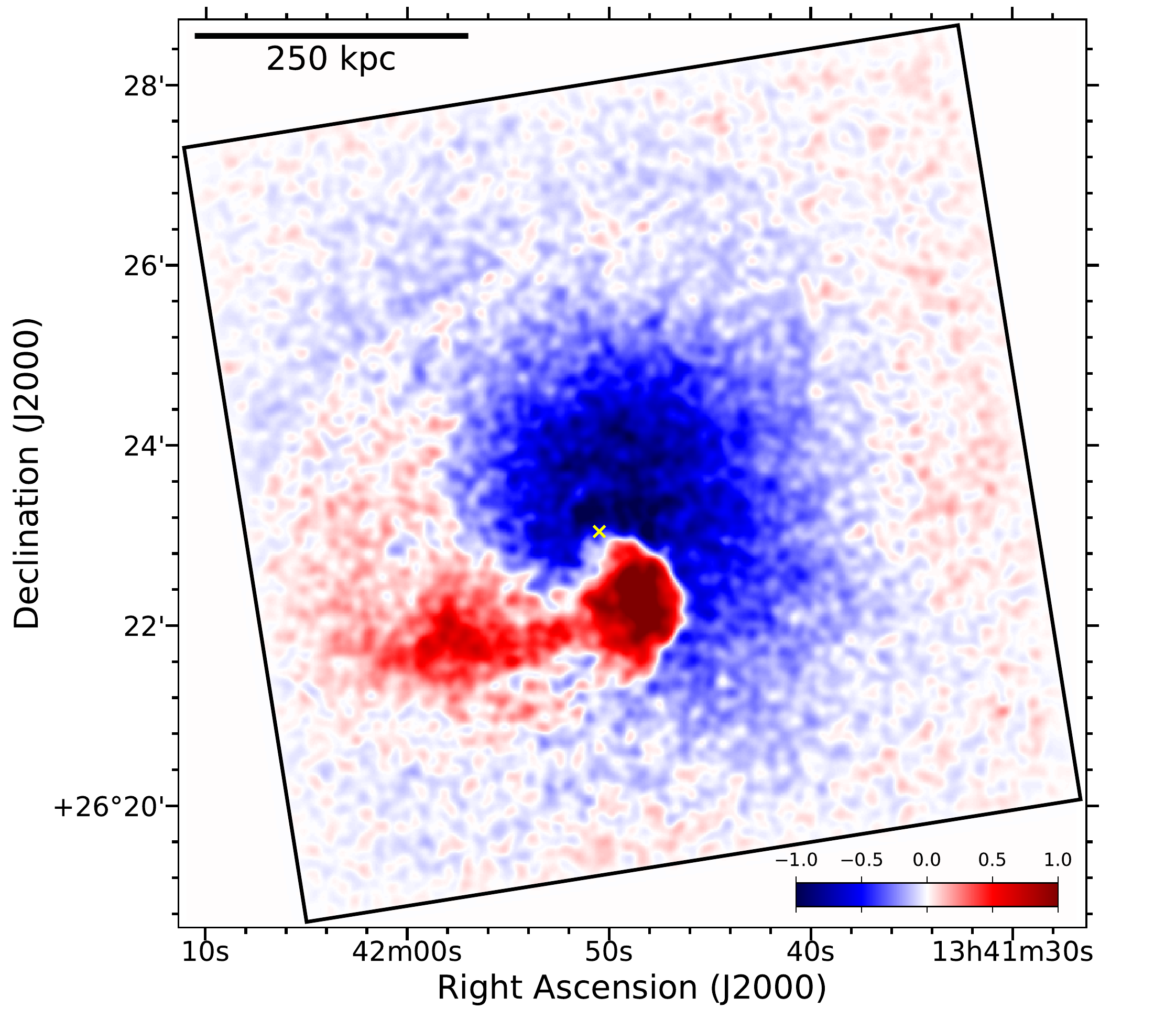}
 \includegraphics[width=.49\hsize,trim={0cm 0cm 0cm 0cm},clip]{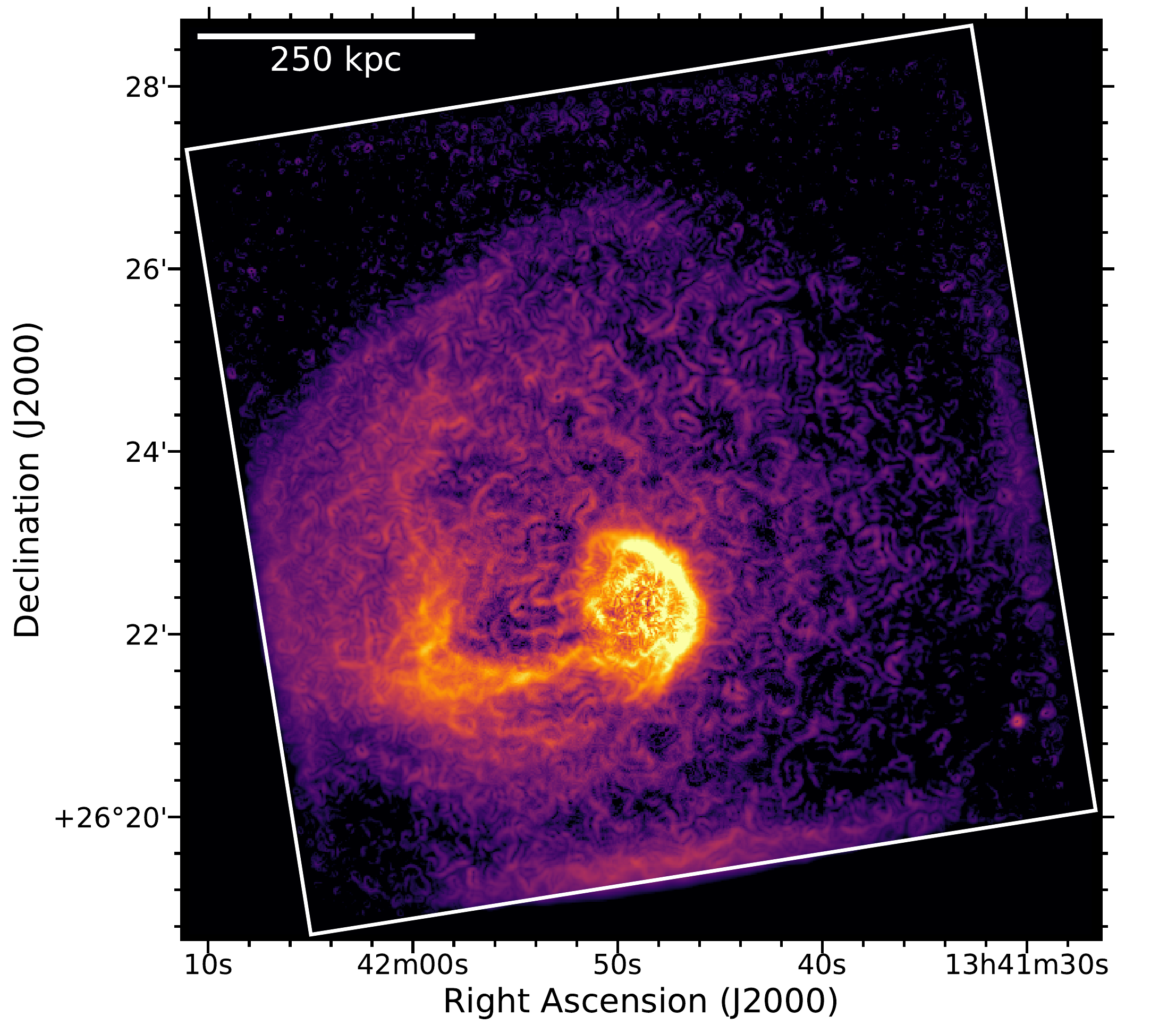}
  \caption{\textit{Left}: Residuals (in units of $\sigma$) between the data and the best-fit $\beta$-model profile centered on the image centroid (marked by the cross). \textit{Right}: GGM-filtered image with radial weighting.}
 \label{fig:spiral}
\end{figure*}

\begin{figure*}
 \centering
 \includegraphics[width=.33\hsize,trim={1.5cm 0cm 1.5cm 0cm},clip]{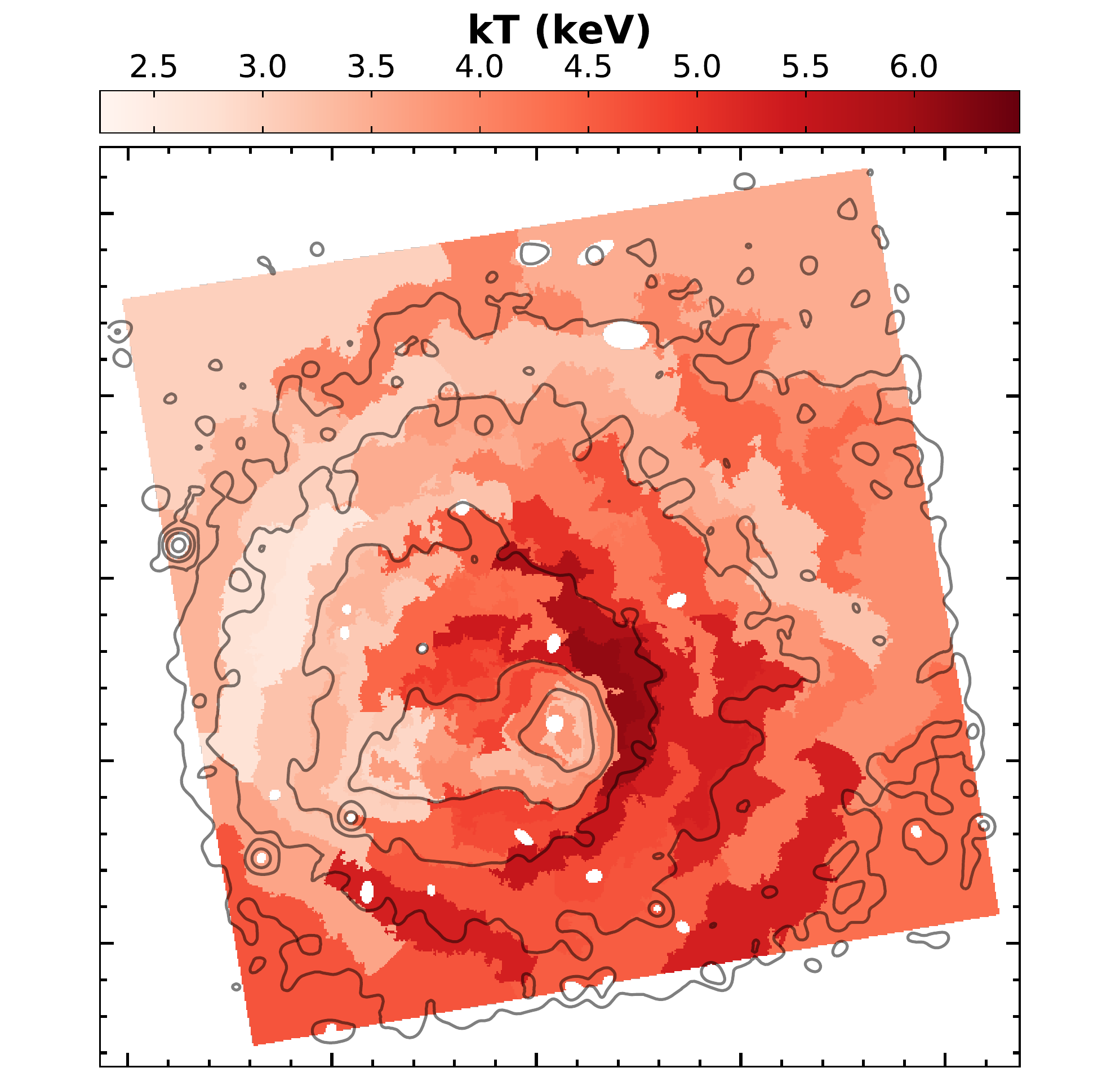}
 \includegraphics[width=.33\hsize,trim={1.5cm 0cm 1.5cm 0cm},clip]{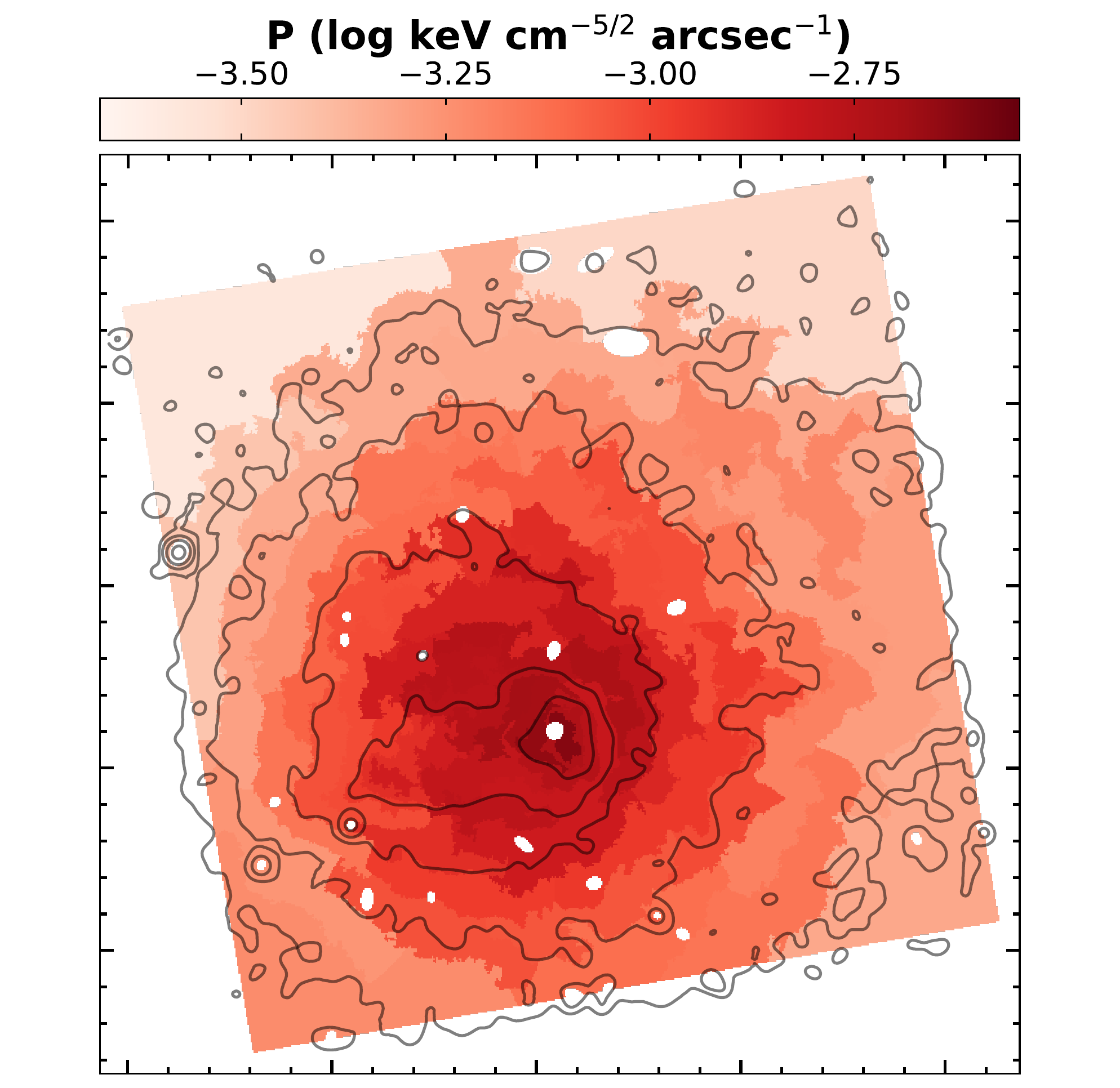}
 \includegraphics[width=.33\hsize,trim={1.5cm 0cm 1.5cm 0cm},clip]{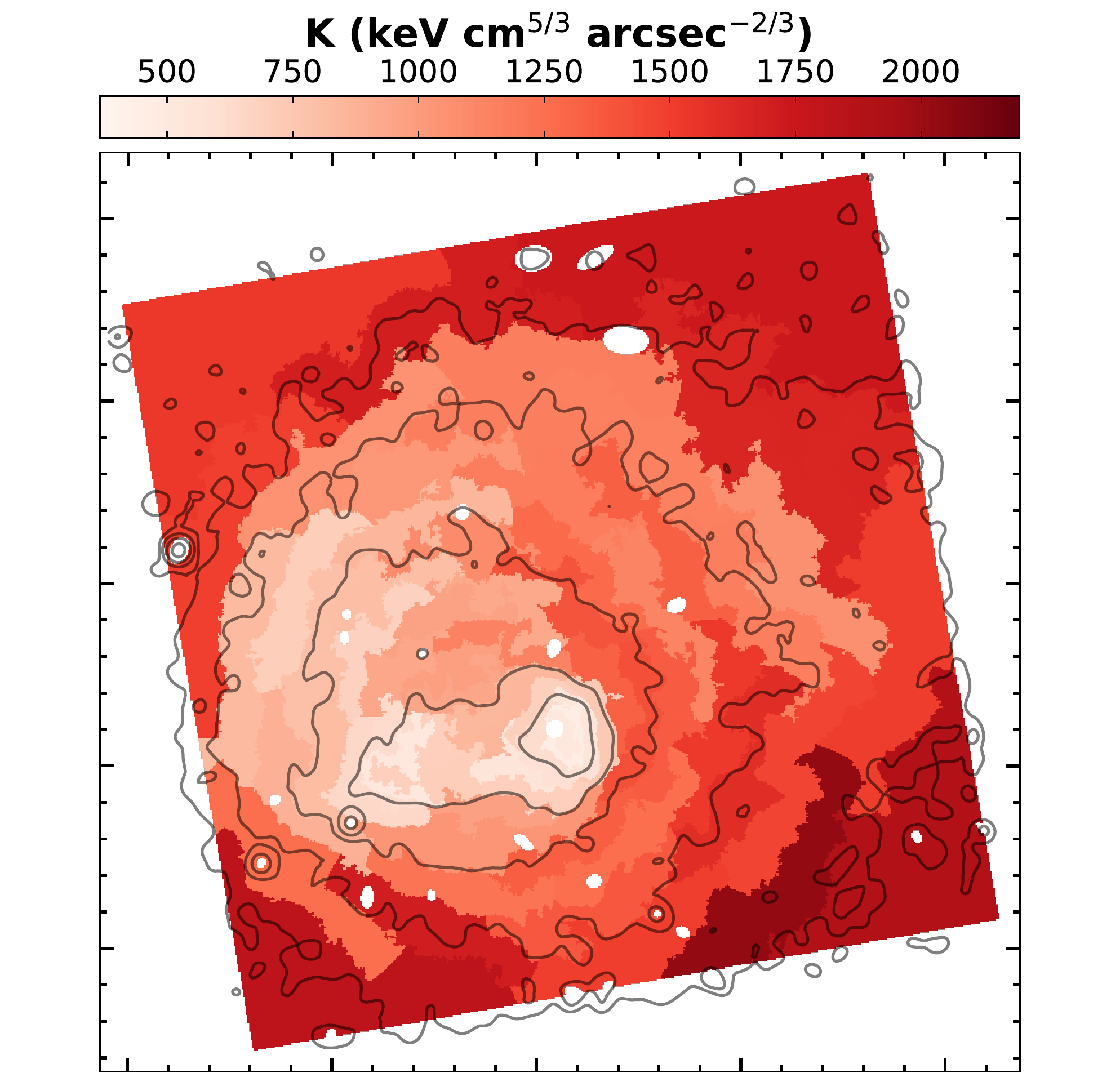}
  \caption{Temperature (\textit{left}), pseudo-pressure (\textit{center}), and pseudo-entropy (\textit{right}) maps with \chandra\ contours overlaid. The white ellipses mark the regions where point sources were removed. The typical fractional error in each region is in the range $\sim 5-10\%$. The temperature error map is shown in Appendix~\ref{app:error_map}.}
 \label{fig:icm_maps}
\end{figure*}

\begin{figure*}
 \centering
 \includegraphics[width=.49\hsize,trim={0cm 0cm 0cm 0cm},clip]{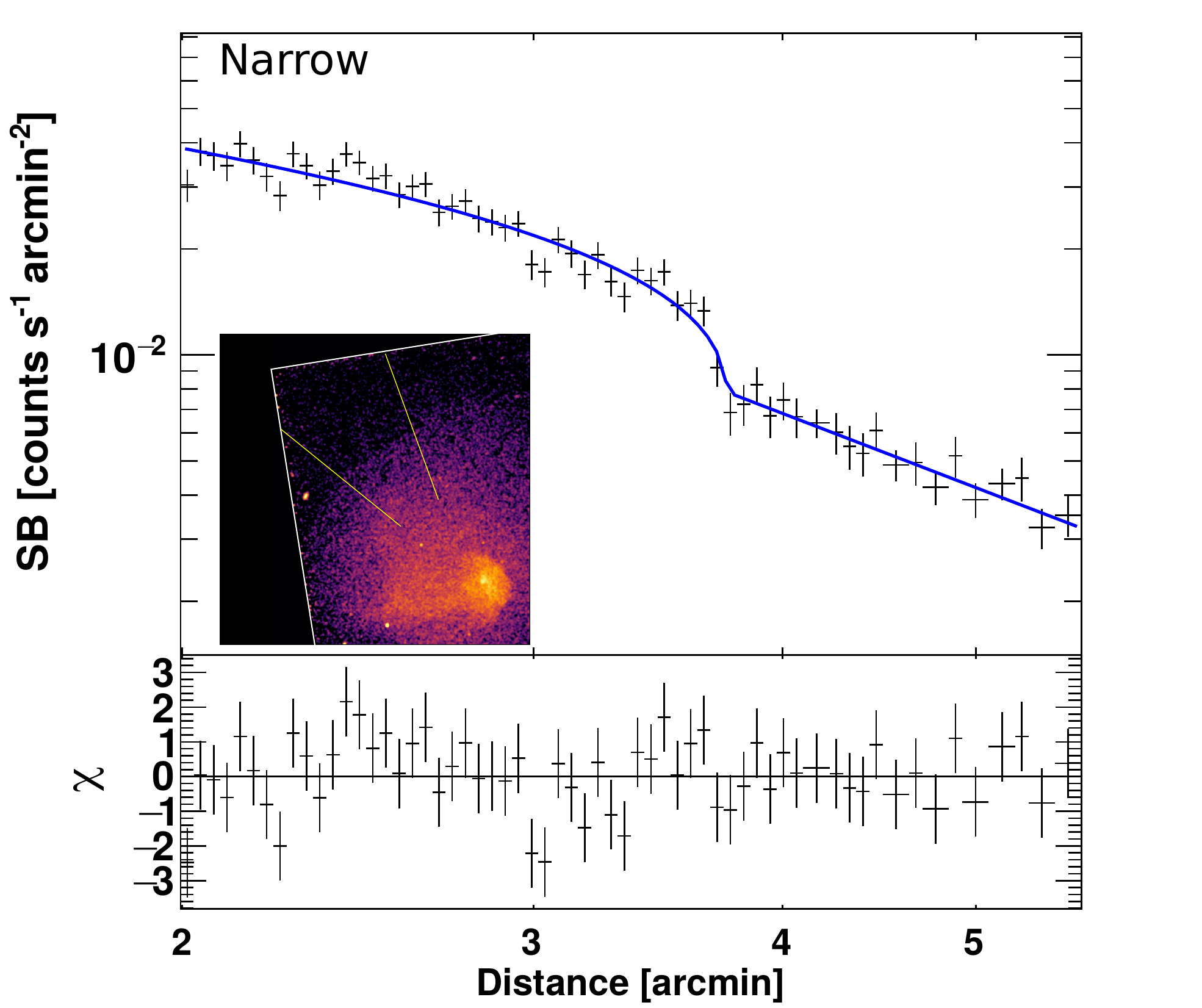}
 \includegraphics[width=.49\hsize,trim={0cm 0cm 0cm 0cm},clip]{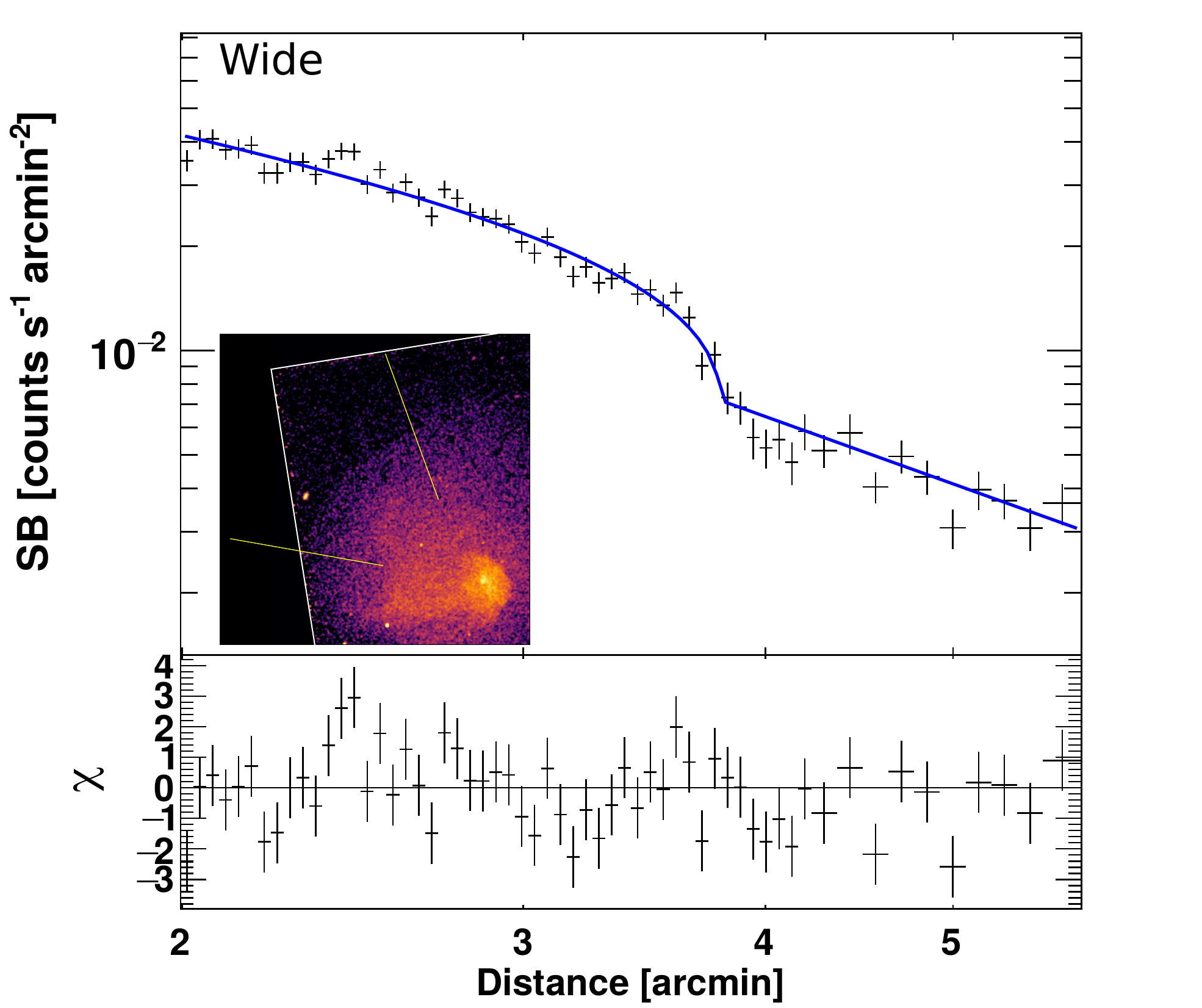}
  \caption{Surface-brightness profiles in the $0.5-2.0$ keV band and best-fit broken power-law models for the NE discontinuity. The inset panels show the sectors used for the analysis. The $\chi^2$/d.o.f. of the fits are  $65.31/56$ (narrow sector) and $73.99/53$ (wide sector).}
 \label{fig:NE_edge}
\end{figure*}

\begin{figure}
 \centering
 \includegraphics[width=\hsize,trim={0cm 0cm 0cm 0cm},clip]{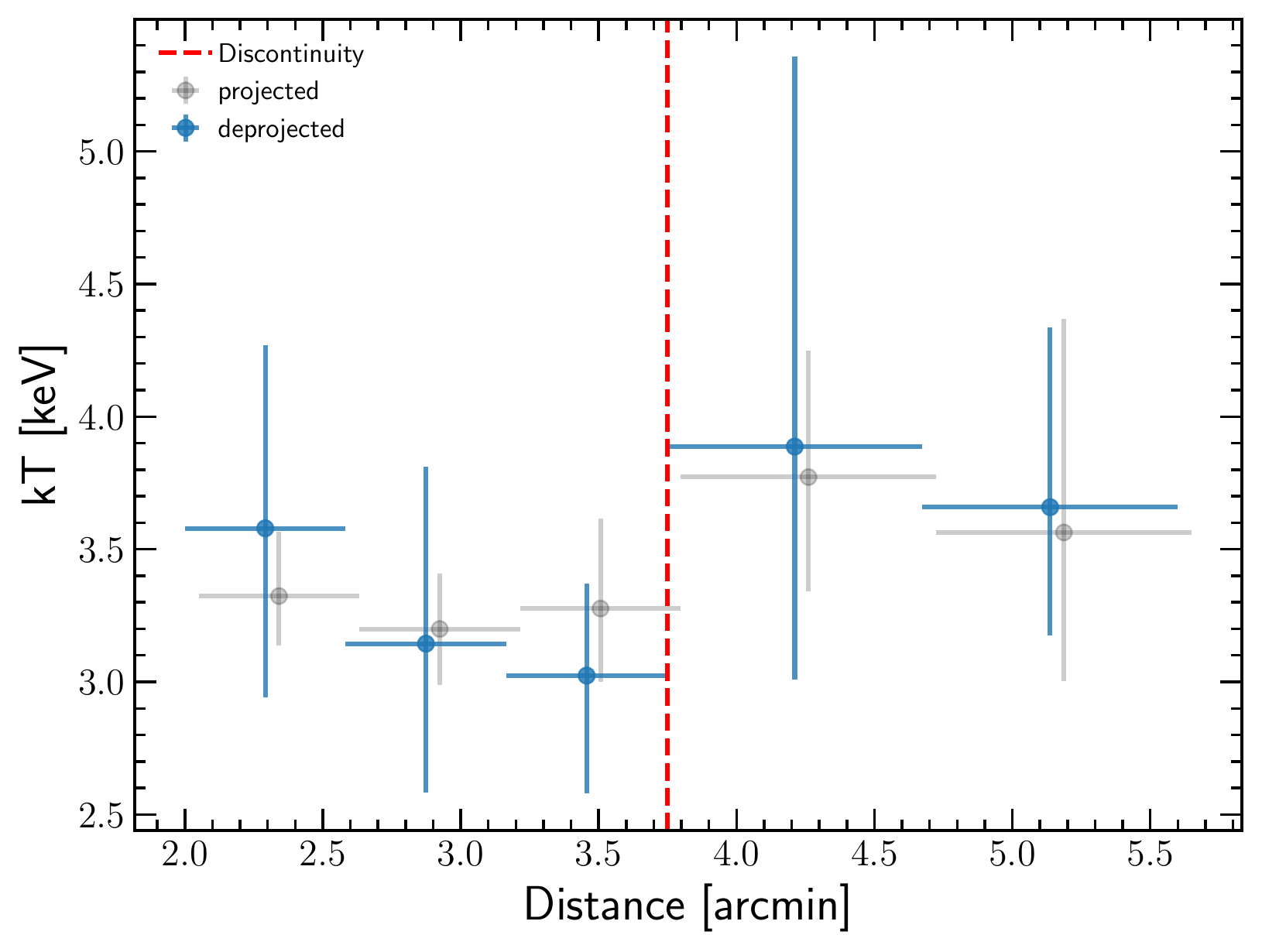}
  \caption{Deprojected (with PROJCT) temperature profile across the narrow sector of Fig.~\ref{fig:NE_edge}. The position of the surface-brightness discontinuity is marked by the dashed line. The projected profile is slightly offset in the $x$-direction for clarity.}
 \label{fig:T_profile}
\end{figure}

\begin{figure*}
 \centering
 \includegraphics[width=.75\hsize,trim={0cm 0cm 0cm 0cm},clip]{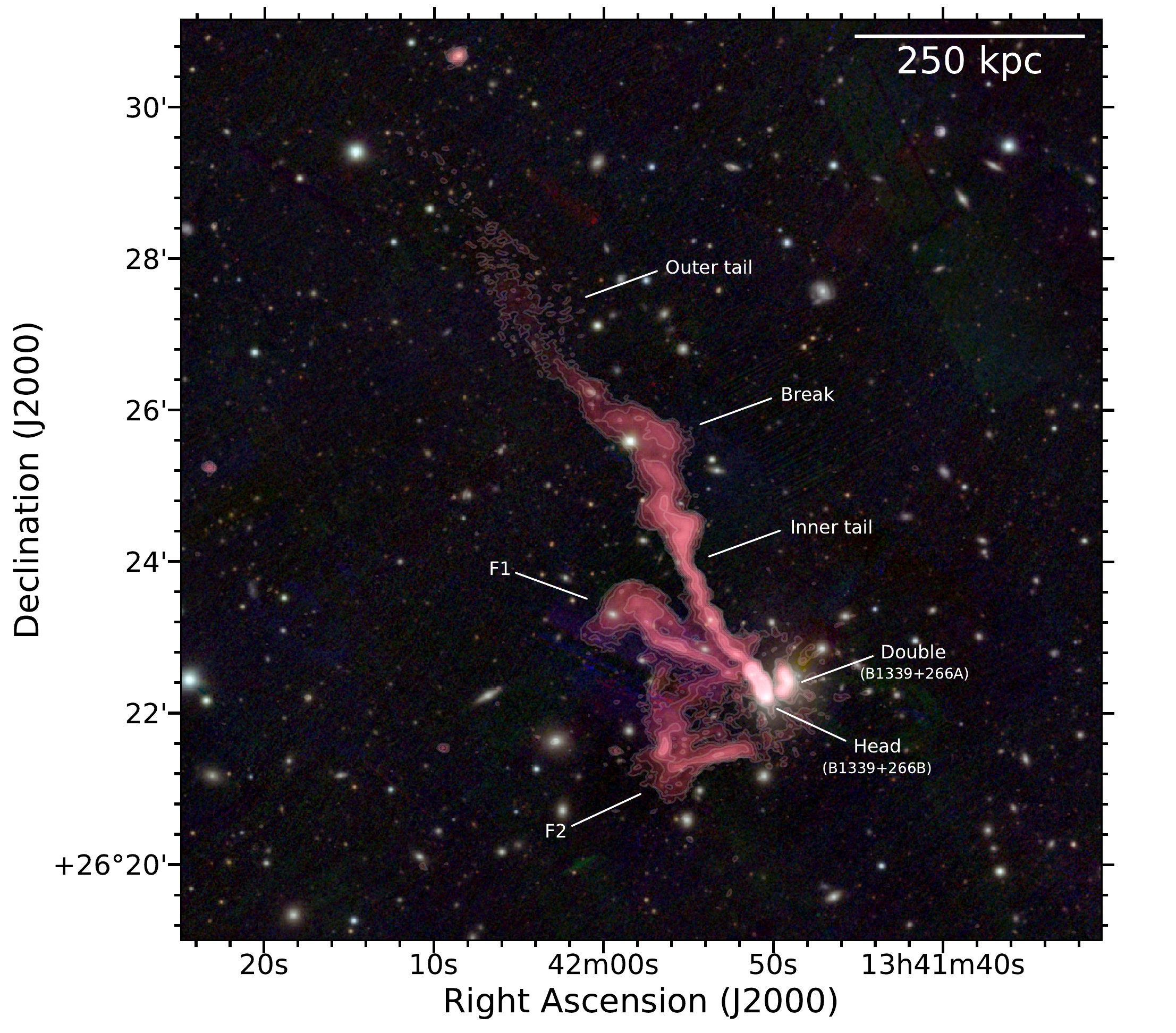}
  \caption{Composite optical-radio image obtained from \panstarrs\ \textit{g,r,i} and \lofar\ 144~MHz high-resolution ($5\arcsec \times 3\arcsec$) data. Radio contours start from $3\sigma$, where $\sigma=148$ \mujyb, and they are spaced by a factor of 2. The main features discussed in the text are labeled in the figure.}
 \label{fig:optical_lofar}
\end{figure*}

A1775 was observed twice (\obsid s: 12891, 13510) with \chandra\ \aciss\ in \vfaint\ mode for a total exposure time of 100~ks. Observations were retrieved from the \chandra\ data archive and processed from the \texttt{level=1} event file with \ciao\ v4.10 and using \caldb\ v4.7.8. We inspected the light curves extracted in the $0.5-7.0$ keV band from the S1 chip with the \texttt{lc\_clean} routine and did not find any periods affected by high background. The two \obsid s were combined with \texttt{merge\_obs} to produce an image in the $0.5-2.0$ keV band binned to have a pixel size of 0.984\arcsec. Point sources were automatically detected with \texttt{wavdetect}, confirmed by visual inspection, and excluded in the subsequent analysis. \\
\indent
We obtained a background image by reprojecting the background templates normalized by counts in the $9.5-12$ keV band to the corresponding event files of the single \obsid s. Surface-brightness analysis was performed with \proffit\ v1.5 \citep{eckert11} on the exposure-corrected and background-subtracted image of the cluster. \\
\indent
For each \obsid, spectra were extracted from the same regions with \texttt{specextract} and jointly fitted with \xspec\ v12.10.0c \citep{arnaud96}. The ICM emission was modeled assuming an absorbed thermal model (PHABS*APEC) with fixed redshift ($z=0.072$) and column density ($\nh = 1.07 \times 10^{20}$ cm$^2$; \citealt{willingale13}). Spectra were fit in the $0.5-7.0$ keV band adopting the solar abundance ratios from \citet{asplund09rev} and using normalized blank-sky observations to create background spectra. Maps of the thermodynamical quantities of the ICM were computed following \citet{botteon18edges} and employing \contbin\ v1.6 \citep{sanders06contbin} to bin a $0.5-7.0$ keV band image adopting a signal-to-noise ratio (S/N) of 50 for the net counts per spectral region.

\subsection{\lofar}

A1775 falls within 2.5\deg\ from the center of four pointings (P207+25, P204+25, P207+27, P204+27) of the \lotssE\ \citep[\lotss;][]{shimwell17, shimwell19}. Each \lotss\ pointing is observed with the \lofar\ High Band Antenna (HBA) array operating at $120-168$ MHz and has an integration time of eight~hours book-ended by ten-minute flux calibrator scans. \\
\indent
In the first step of the data reduction, the four observations were analyzed individually with the pipelines developed by the \lofar\ Surveys Key Science Project team (\prefactor, \citealt{vanweeren16calibration, williams16, degasperin19}; \killms, \citealt{tasse14arx, tasse14, smirnov15}; \ddfacet, \citealt{tasse18}) to correct for direction-independent and dependent effects. In particular, we deployed the calibration pipeline\footnote{\url{https://github.com/mhardcastle/ddf-pipeline}} v2.2, which includes improvements in terms of dynamic range, deconvolution of diffuse emission, image artifacts and image fidelity in general, delivering $6\arcsec$ resolution images of the entire \lofar\ field-of-view (FoV) with noise levels $\sim 100$ \mujyb\ (see Section~5 in \citealt{shimwell19} and \citealt{tasse21}). \\
\indent
In the second step, we subtracted the sources outside a $33\arcmin\times33\arcmin$ region that contains A1775 from the visibility data of each pointing by using the models derived from the pipeline. The extracted datasets are then phase-shifted to the center of the region, averaged, and corrected for the \lofar\ station beam toward this direction. Residual artifacts were attenuated by means of phase and amplitude self-calibration loops on the combined measurement sets. This procedure is discussed in detail in \citet{vanweeren20arx} and has already been adopted in several \lofar\ HBA works \citep[\eg,][]{hardcastle19, cassano19, botteon19lyra, botteon20a2255, osinga21, hoang21a990}. \\
\indent
After the self-calibration step, the four observations are jointly imaged and deconvolved with \wsclean\ v2.8 \citep{offringa14}. Final images at the central observing frequency of 144 MHz were obtained using the multiscale multifrequency deconvolution scheme \citep{offringa17} with Briggs weightings \citep{briggs95} \texttt{robust=-1.25} for the highest resolution image, or \texttt{robust=-0.5} for lower resolution images. \\
\indent
It is known that the \lofar\ flux density scale can show systematic offsets \citep[\eg,][]{hardcastle16}. For this reason, we compared the flux density of the brightest compact sources in our new images with the mosaiced images from the forthcoming second \lotss\ data release (Shimwell et al., in prep.). Therefore, we adopted a correction factor of 0.96 on \lofar\ flux densities derived from the mean flux density ratio between our image and the \lotss-DR2 mosaic. We set a conservative systematic uncertainty of 20\% on \lofar\ flux density measurements, as was done by \lotss\ \citep{shimwell19}.

\subsection{\gmrt}

We reanalyzed \gmrt\ observations of A1775 at 235~MHz and 610~MHz that were previously published in \citet{giacintucci07}. The cluster was observed in 2005 for 100 minutes at 235~MHz and in 2003 for 45 minutes at 610~MHz using the old hardware correlator and a bandwidth of 8 MHz and 16 MHz, respectively. We refer the reader to \citet{giacintucci07} for further details
on these observations. \\
\indent
We reprocessed the 235 MHz data using the \aipsE\ (\aips) starting from the raw visibilities to take advantage of the radio frequency interference (RFI) excision task \texttt{RFLAG,} which makes it possible to perform a careful flagging of bad data and was not available at the time of the previous analysis. We reduced the data following standard procedure, using 3C286 and 3C147 as primary calibrators to compute and apply gain and bandpass corrections to the data. 3C286 was also used to calibrate the data in phase. We used a combination of \texttt{RFLAG} and manual flagging on each source individually (calibrators and target) to remove RFI-affected visibilites and other bad data. We then appropriately averaged the calibrated target data to 21 frequency channels, each 250 KHz wide, to reduce the dataset size while avoiding significant time or bandwidth smearing. We applied a number of phase-only self-calibration cycles to the target visibilities and used wide-field imaging by decomposing the primary beam area into $\sim40$ smaller facets. We produced final images with the multi-scale \texttt{CLEAN} implemented in \texttt{IMAGR}. The rms sensitivity level achieved in the image at full resolution ($14\arcsec\times10\arcsec$, for \texttt{ROBUST=0} in \texttt{IMAGR}) is $\sim 0.8$ mJy beam$^{-1}$, that is, $\sim 50\%$ lower than the image noise in \citet{giacintucci07}. We also produced images with lower resolution (down to $\sim 28\arcsec$) by increasing the \texttt{ROBUST} parameter and/or applying tapers to the \emph{\emph{\emph{\textit{\uv}}}} data during the imaging process. Overall, the combination of lower noise and better \uv\  coverage (resulting from the accurate RFI excision done with \texttt{RFLAG}) makes these new images more sensitive to faint and diffuse emission than that of \citet{giacintucci07}.  \\
\indent
We also attempted to recalibrate the 610~MHz data presented in \citet{giacintucci07}; however, no improvement was achieved compared the quality obtained in the previous work (we note thet RFI is less critical at this frequency than at 235 MHz). Therefore, we used the calibrated data at 610~MHz from \citet{giacintucci07} to produce a new image at a resolution of $15\arcsec$ using the multi-scale \texttt{CLEAN}. \\
\indent
We corrected all images for the the \gmrt\ primary beam response\footnote{\url{http://www.ncra.tifr.res.in:8081/~ngk/primarybeam/beam.html}} using \texttt{PBCOR} in \aips. Residual amplitude errors are estimated to be within $15\%$ at 235 MHz and $10\%$ at 610 MHz \citep[\eg,][]{chandra04}.

\subsection{\vla}

We obtained observations of A1775 at 1.4~GHz from the \vla\ public archive. The cluster was observed in 2001 using the D-array configuration for a total of 1.4 hour (project AL0515) and in 2004 for 2.5 hour in C-configuration (project AV0265). We calibrated the data from the two configurations separately using \aips\ and standard amplitude and phase calibration procedures. Bandpass calibration was applied to the D-configuration data that were taken using eight frequency channels in each of the two 25 MHz-wide intermediate frequency (IF) subbands. The flux density scale was set using the primary calibrators 3C147 and 3C286 and the \citet{perley17} coefficients in \texttt{SETJY}. We applied several loops of phase self-calibration to each dataset to reduce the effects of residual phase errors in the data. We then averaged the D-configuration visibilities to one single channel/IF and combined them with the C-configuration data in the \uv\  plane. Finally, we applied a final step of self-calibration in phase to the combined dataset. Residual flux calibration errors are estimated to be within 5\%. %

\section{Results}

\subsection{X-ray arc-shaped feature}\label{sec:arc-shaped}

The deep \chandra\ observation reveals an unusual system characterized by a bright core bounded by a sharp and prominent front in the W creating a mushroom-cap structure (Fig.~\ref{fig:chandra_image}). The bulk of the X-ray emission seems offset toward the NE, where a clear drop in surface brightness is also observed. To enhance the presence of these features, we produced the images shown in Fig.~\ref{fig:spiral}. On the left-hand panel, we report the residuals between the data and the best-fit spherical $\beta$-model\footnote{In Appendix~\ref{app:elliptical_model}, we show that adopting an elliptical $\beta$-model does not lead to any appreciable difference.} \citep{cavaliere76} centered on the image centroid (at RA: 13d41m51s, DEC: +26\deg23\arcmin00\arcsec) that was adopted to describe the X-ray surface brightness (core radius $r_{\rm c} = 2.56\pm0.05$ arcmin and $\beta = 0.84 \pm 0.01$). This image highlights a spiral pattern extending in clockwise direction from the cluster core. On the right-hand panel, we report the Gaussian gradient magnitude (GGM) image \citep{sanders16ggm}, which was obtained by combining images filtered on scales $\sigma_{\rm GGM} = [1,2,4,8,16]$ pixels with radial weighting. This representation highlights the gradients of an image, such as surface-brightness discontinuities. For A1775 it emphasizes the leading edge ahead of the cluster core (the mushroom with a stem and a cap), as well as the same spiral noted in the residual image. \\
\indent
We investigated the thermodynamical properties of the cluster with temperature, pseudo-pressure, and pseudo-entropy maps of ICM that are produced in the way described in Section~\ref{sec:chandra}. These maps are shown in Fig.~\ref{fig:icm_maps}. In the mushroom, we detect a cool ($\sim 3$ keV) X-ray tail and a temperature jump at its W surface-brightness front. The temperature is lower closer to the cluster center, indicating that we are observing a cold front. A detailed study of this edge will be presented in a forthcoming paper. Once again, a clear spiral pattern is observed in temperature and pseudo-entropy, where low-temperature and low-entropy gas is found to trace the overdense region of Fig.~\ref{fig:spiral}. In the NE direction, the \contbin\ algorithm found only one spectral region satisfying the requested S/N because of lower count statistics due to the drop in surface brightness and greatest off-axis distance. A detailed analysis of this region is presented in the following. \\
\indent
At $\sim 500$ kpc from the cluster core, toward the NE, an arc-sharped discontinuity can be seen by eye (Fig.~\ref{fig:chandra_image}). The feature is even clearer in the residual map and GGM-filtered image of Fig.~\ref{fig:spiral}, as well as by the spectral regions of Fig.~\ref{fig:icm_maps} drawn by \contbin, which are optimized to follow the cluster X-ray brightness. We quantify the density jump in this direction by extracting and fitting X-ray surface-brightness profiles assuming spherical symmetry and an underlying a broken power-law density profile in the following form:

\begin{equation}\label{eq:bknpow}
 \begin{array}{ll}
 n_{\rm in} (r) = \compr n_0 \left( \frac{r}{r_{\rm j}} \right)^{a_1}, & \mbox{if} \quad r \leq r_{\rm j} \\
\\
 n_{\rm out} (r) = n_0 \left( \frac{r}{r_{\rm j}} \right)^{a_2}, & \mbox{if} \quad r > r_{\rm j}
 \end{array}
,\end{equation}

\noindent
where $r$ denotes the radius from the center of the sector, $r_{\rm j}$ is the radius of the jump, $\compr \equiv n_{\rm in}/n_{\rm out}$ is the density ratio, $n_{\rm in}$ and $n_{\rm out}$ are the inner and outer densities, $n_0$ is a normalization factor, while $a_1$ and $a_2$ are the power-law indices. This choice is customarily adopted to model discontinuities in the ICM, such as shocks and cold fronts \citep[\eg,][]{markevitch07rev}. The fitting was done leaving all the parameters free to vary. The analysis was performed in a ``narrow'' and a ``wide'' sector (centered at RA: 13d41m51s, DEC +26\deg22\arcmin24\arcsec). The former is entirely covered by the S3 chip and its angular aperture toward the W is extended up to the position of a surface-brightness kink. The latter covers the whole angular length of the discontinuity and extends beyond the chip FoV. The best-fit models and regions of the analysis are shown in Fig.~\ref{fig:NE_edge}. We find that the broken-power law model provides a good description of the data. The density ratio (deprojected along the line of sight under the assumption of spherical geometry by \proffit) and radial distance of the discontinuity are consistent between the two regions: $\compr = 1.67^{+0.11}_{-0.10}$ and $r_{\rm j} = 3.75^{+0.02}_{-0.03}$ arcmin (narrow sector) and $\compr = 1.73^{+0.11}_{-0.10}$ and $r_{\rm j} = 3.79^{+0.03}_{-0.02}$ arcmin (wide sector). The best-fit radius of the jump is slightly offset in the wide sector due to the presence of a kink in the discontinuity (at RA: 13h42m03s, DEC: $+26\deg24\arcmin57\arcsec$, see Fig.~\ref{fig:chandra_image}), possibly suggesting that its curvature radius mildly changes across the front. \\
\indent
Spectral analysis across the discontinuity was performed in five regions (3 inwards and 2 outwards) obtained from the narrow sector of Fig.~\ref{fig:NE_edge}, which avoids the edge of the S3 chip. In order to take into account the contribution of each annular shell into the neighboring ones, we made use of the PROJCT model in \xspec\ to deproject the data in 3D under the assumption of spherical symmetry. In Fig.~\ref{fig:T_profile}, we report the projected and deprojected radial temperature profiles. Although the large uncertainties do not allow us to claim a sharp temperature jump, we find that the internal region is characterized by lower temperature values, as observed in the case of cold fronts. In the two spectral sectors just inwards and outwards of the surface-brightness edge, we measure $kT_{\rm in} = 3.0^{+0.3}_{-0.4}$ keV and $kT_{\rm out} = 3.9^{+1.5}_{-0.9}$ keV, respectively. The observation of an arc-shaped cold front accompanied by a spiral pattern in the residual and gradient images and temperature and pseudo-entropy maps, indicate the presence of substantial gas motions in the central region of A1775. The dynamical state of the cluster and the origin of the arc-shaped feature are discussed in Section~\ref{sec:dyn-state}.

\begin{figure*}
 \centering
 \includegraphics[width=.8\hsize,trim={0cm 0cm 0cm 0cm},clip]{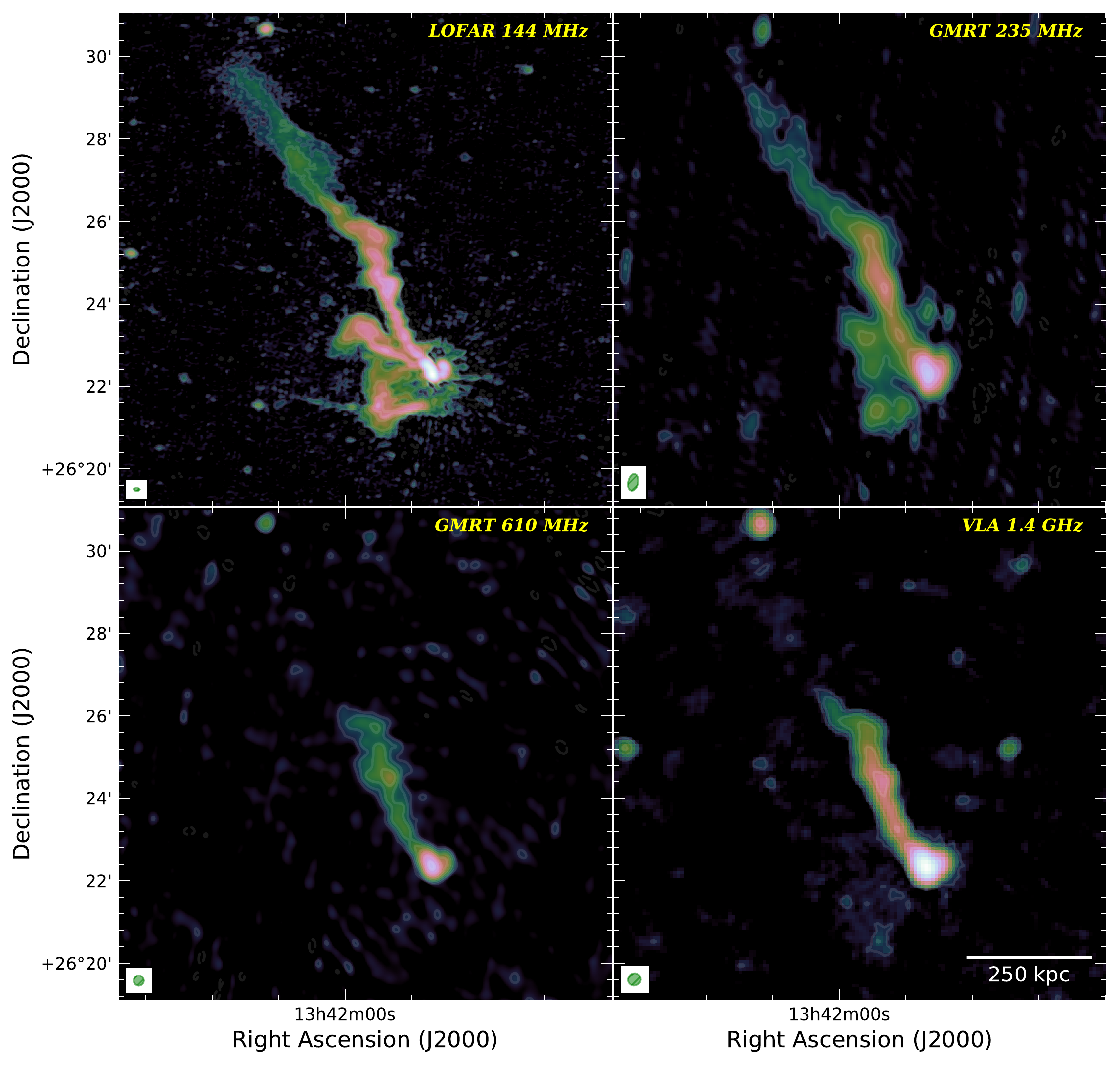}
 \caption{\lofar\ 144~MHz ($9\arcsec \times 5\arcsec$), \gmrt\ 235~MHz ($26\arcsec \times 14\arcsec$), \gmrt\ 610~MHz ($15\arcsec \times 15\arcsec$), and \vla\ C+D array 1.4~GHz ($19\arcsec \times 18\arcsec$) images. Contours are spaced by a factor of 2 starting from $3\sigma$, where $\sigma_{144} = 105$ \mujyb, $\sigma_{235} = 0.9$ \mjyb, $\sigma_{610} = 680$ \mujyb, and $\sigma_{1400} = 66$ \mujyb. The color scale has a logarithmic stretch from 1 to $3000\sigma$. Beams are shown in the bottom left corners.}
 \label{fig:radio_images}
\end{figure*}

\subsection{Head-tail radio galaxy}\label{sec:results-ht}

\begin{figure}
 \centering
 \includegraphics[width=\hsize,trim={0cm 0cm 0cm 0cm},clip]{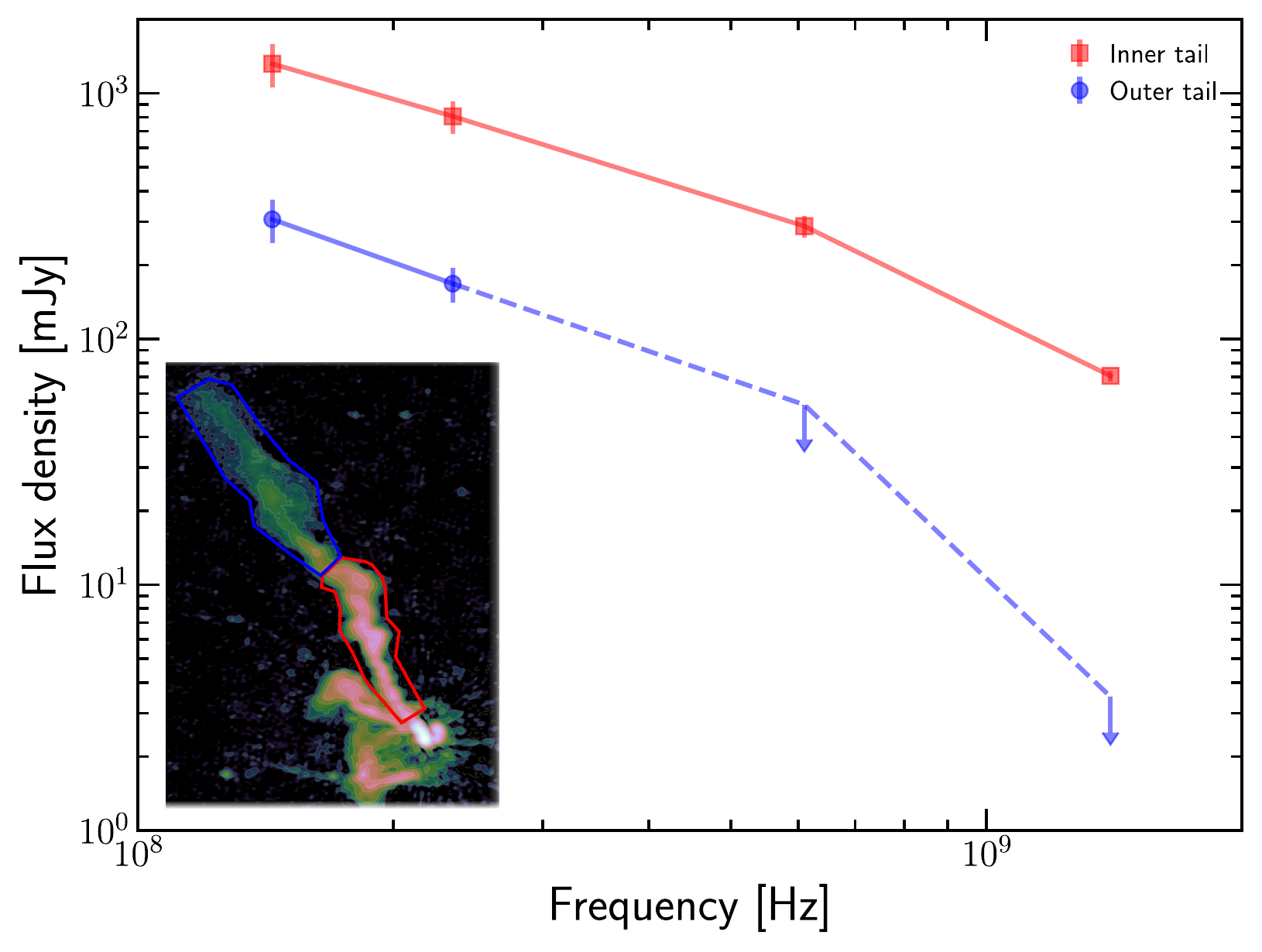} 
 \caption{Integrated spectra of the ``inner'' and ``outer'' regions (shown in the inset panel) of the head-tail radio galaxy.}
 \label{fig:spectra_tail}
\end{figure}

The most prominent radio sources in A1775 are the extended head-tail radio galaxy (B1339+266B) and the double radio source (B1339+266A), which originate from two elliptical galaxies of similar luminosity, separated by a projected distance of 32 kpc \citep{parma91}. The association between the radio emission and the host galaxy was done by \citet{giacintucci07}. In Fig.~\ref{fig:optical_lofar}, we show our \lofar\ 144 MHz image of A1775 at a resolution of $5\arcsec \times 3\arcsec$ overlaid with an optical image from the \panstarrsE\ \citep[\panstarrs;][]{chambers16arx}. Both sources are detected at high S/N, and some residual artifacts due to ionospheric phase corruptions can be noticed around the two elliptical galaxies. The radio structures in the eastern direction are discussed in the next Section. \\
\indent
A multifrequency view of the radio emission from A1775 at 144~MHz (\lofar), 235~MHz (\gmrt), 610~MHz (\gmrt),  and 1.4~GHz (\vla) is shown in Fig.~\ref{fig:radio_images}. The \gmrt\ and \vla\ images are shown at lower resolution to enhance the sensitivity to extended emission, which is better recovered in the \lofar\ image despite the higher resolution. The emission of the head-tail radio galaxy can roughly be divided into three components: the bright head, the previously reported emission of the tail extending for $\sim400$ kpc, and a newly discovered low-surface-brightness extension that is only observed at 144~MHz and 235~MHz (Fig.~\ref{fig:radio_images}). The latter component is diffuse and originates after a ``break'' of the collimated tail. The newly discovered diffuse emission from the tail extends toward the NE for an additional $\sim400$ kpc and is not detected at higher frequencies, likely due to its steep spectrum. Overall, the head-tail radio galaxy in A1775 spans $\sim800$ kpc (projected), making it one of the longest tailed sources observed in galaxy clusters to date (see \eg,\ Abell 1132 \citep{wilber18a1132} and Abell 2256 \citep{owers14} for two noteworthy examples). From our \lofar\ high-resolution image that is shown in Fig.~\ref{fig:optical_lofar}, we measure a total flux density of $S_{144} = 2.8\pm0.6$ Jy for the tailed source (uncertainties are dominated by the 20\% calibration error). This flux density is distributed among the three components of the head-tail radio galaxy as follows: 1.2 Jy for the head, 1.3 Jy for the ``inner'' 400~kpc-tail, and 0.3 Jy for the ``outer'' tail extension. In Fig.~\ref{fig:spectra_tail}, we show the values of the tail together with the \gmrt\ and \vla\ flux densities, that are also summarized in Tab.~\ref{tab:fluxes} together with the other radio sources in A1775. Since the outer tail is not detected at 610~MHz and 1.4~GHz, we report the $1\sigma$ upper limits on the emission. The inner tail follows a power-law with $\alpha^{610}_{144} = 1.06\pm0.02$ between 144~MHz and 610~MHz (the 235~MHz data is included in this estimate), then it steepens to $\alpha^{1400}_{610} = 1.69\pm0.14$ at higher frequencies, suggesting spectral curvature. The outer tail spectral index is not well constrained between 144~MHz and 235~MHz due to the narrow frequency span, implying $\alpha^{235}_{144} = 1.23\pm0.52$. We note that this range of values is not consistent with the upper limit at 1.4~GHz if the spectrum were a power law. This indicates that the spectrum of this component is curved. %

\begin{figure}
 \centering
 \includegraphics[width=\hsize,trim={0cm 0cm 0cm 0cm},clip]{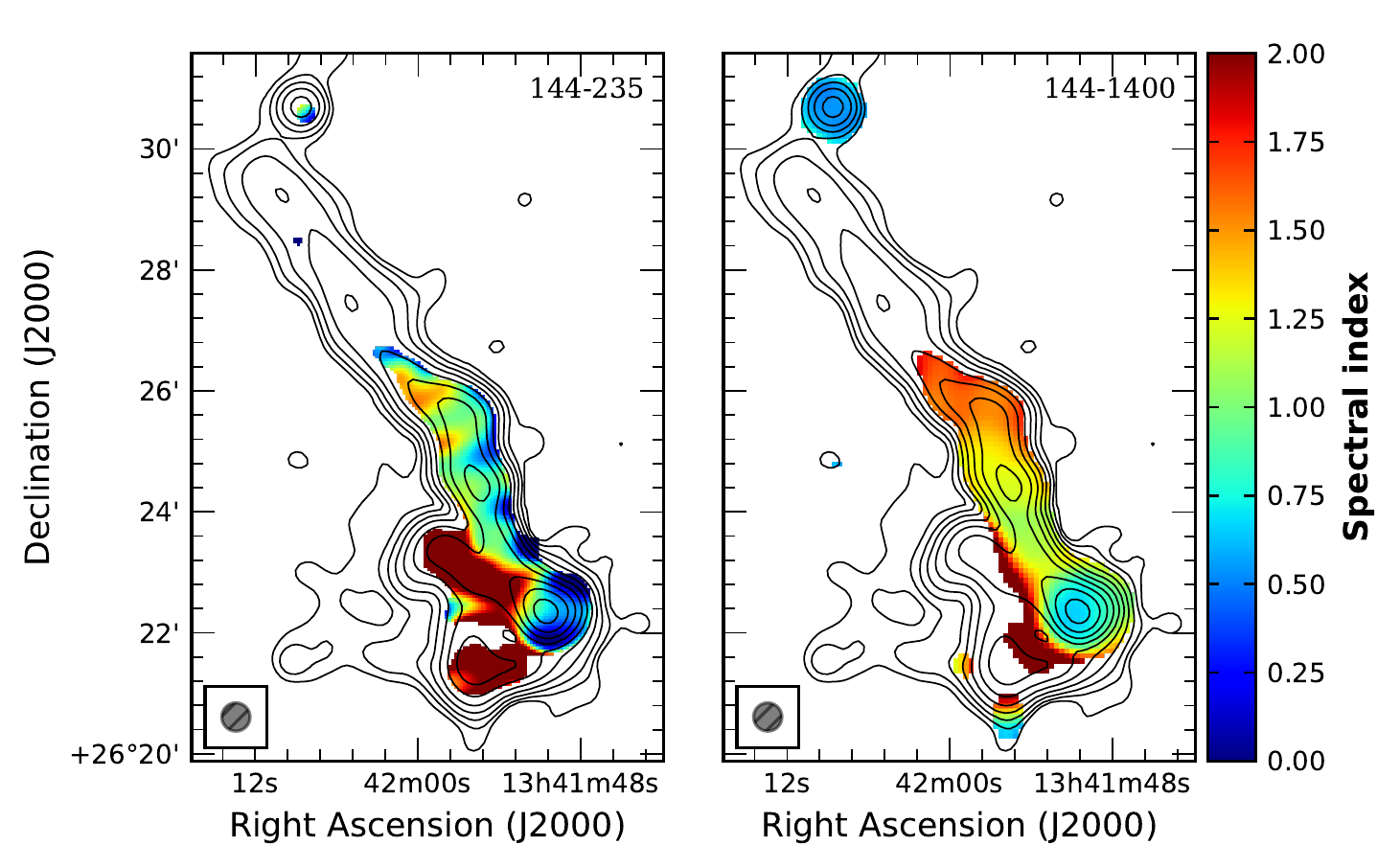} 
 \caption{Low ($144-235$~MHz) and high ($144-1400$~MHz) frequency spectral index maps at a resolution of $28\arcsec\times28\arcsec$ with \lofar\ contours at the same resolution overlaid. The beam is shown in the bottom left corner. Pixels with values below $5$ and $3\sigma$ were blanked in the low- and high-frequency spectral index maps, respectively. The error maps are shown in Appendix~\ref{app:error_map}.}
 \label{fig:spix}
\end{figure}

We used the \gmrt\ 235~MHz and \vla\ 1.4~GHz observations, together with the \lofar\ 144~MHz dataset, to create low- $(144-235$~MHz) and high- $(144-1400$~MHz) frequency spectral index images. To measure the same spatial scales, we applied the inner and outer \uv\  ranges of \gmrt\ and \vla\ observations to our \lofar\ dataset to create the images used to produce the $144-235$~MHz and the $144-1400$~MHz spectral index maps. These ranges were $70\lambda$ to $23$k$\lambda$ and $160\lambda$ to $16.2$k$\lambda$ for the \gmrt\ and \vla, respectively. The images at each frequency band were convolved to the same resolution ($28\arcsec\times28\arcsec$), corrected for positional offsets, and regridded to have an identical pixelation before combination. These spectral index maps, shown in Fig.~\ref{fig:spix}, are useful to study the aging of relativistic electrons in synchrotron emitting sources. In particular, the $144-1400$~MHz map, thanks to its broad coverage that spans one decade in frequency, allows us to provide good constraints on the spectral index of the radiation. However, it is insensitive to the steep spectrum emission detected only at low frequencies. In this respect, the $144-235$~MHz map can give indications about the presence of the steepest spectrum emission, with the caveat that the precise value of $\alpha$ can be affected by the uncertainties introduced by the narrow frequency span. We note that both the maps in Fig.~\ref{fig:spix} show typical values in the core of $\alpha=0.6-0.7$, and indicate spectral steepening along the tail. As mentioned before, the slightly lower (and noiser) values of the $144-235$~MHz spectral index map are likely related to the shorter frequency pivot between the two observations and to the spectral curvature at higher frequencies. In Section~\ref{sec:ht-analysis}, we discuss the surface brightness and spectral features observed along the head-tail radio galaxy in detail.

\begin{table}[t]
 \centering
 \caption{Flux densities of the radio sources in A1775.}
 \label{tab:fluxes}
  \begin{tabular}{llcc} 
  \hline
  \hline
  Source & & $\nu$ [MHz] & $S_\nu$ [mJy]  \\
  \hline
  Double & & 144 & $2402\pm480$  \\
  Head & & 144 & $1213\pm243$  \\
  \multirow{4}*{Inner tail} & \ldelim[{4}{*} \hspace{-0.5cm}  & 144  & $1319\pm264$ \\
                            &                                 & 235  & $806\pm121$ \\ 
                            &                                 & 610  & $288\pm29$ \\ 
                            &                                 & 1400 & $70.1\pm3.6$ \\ 
  \multirow{4}*{Outer tail} & \ldelim[{4}{*} \hspace{-0.5cm}  & 144  & $307\pm61$ \\
                            &                                 & 235  & $168\pm27$ \\ 
                            &                                 & 610  & $<54$ \\ 
                            &                                 & 1400 & $<3.5$ \\
  F1 & & 144 & $593\pm119$ \\
  F2 & & 144 & $505\pm101$ \\
  Central diffuse emission & & 144 & $244\pm50$  \\
  \hline
  \end{tabular}
\end{table}

\subsection{Revived fossil plasma and central diffuse radio emission}

\begin{figure}
 \centering
 \includegraphics[width=\hsize,trim={0cm 0cm 0cm 0cm},clip]{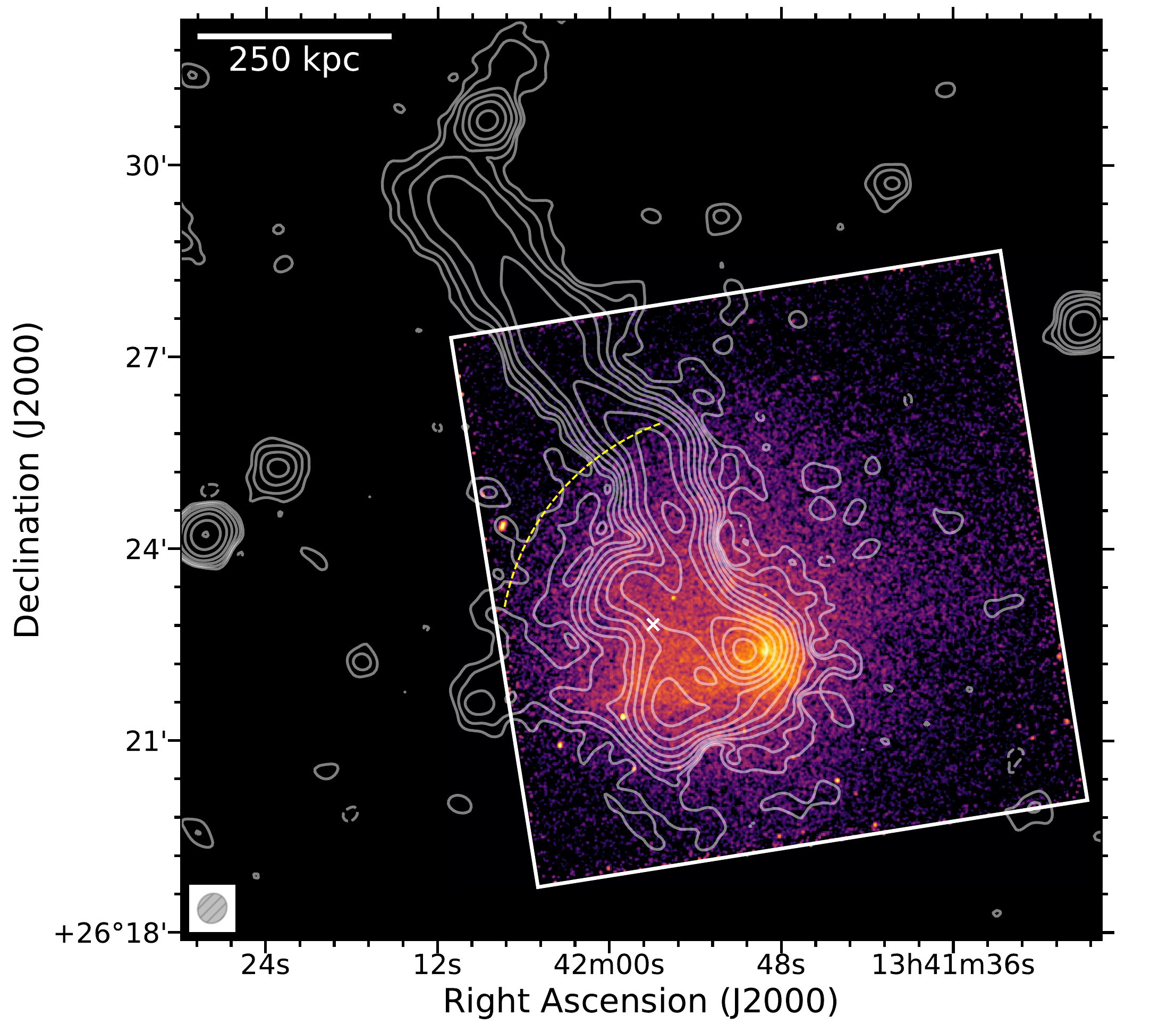}
  \caption{Low-resolution ($29\arcsec\times26\arcsec$) \lofar\ contours overlaid on the \chandra\ image. Contours are spaced by a factor of 2 starting from $3\sigma$, where $\sigma=255$ \mujyb. The beam is shown in the bottom left corner. The dashed yellow arc marks the position of the cold front traced by the X-ray surface-brightness discontinuity of Fig.~\ref{fig:NE_edge} while the white cross denotes the center of the circular model shown in Fig.~\ref{fig:halo_profile}.}
 \label{fig:chandra_lofar}
\end{figure}

\begin{figure*}
 \centering
 \includegraphics[width=\hsize,trim={0cm 0cm 0cm 0cm},clip]{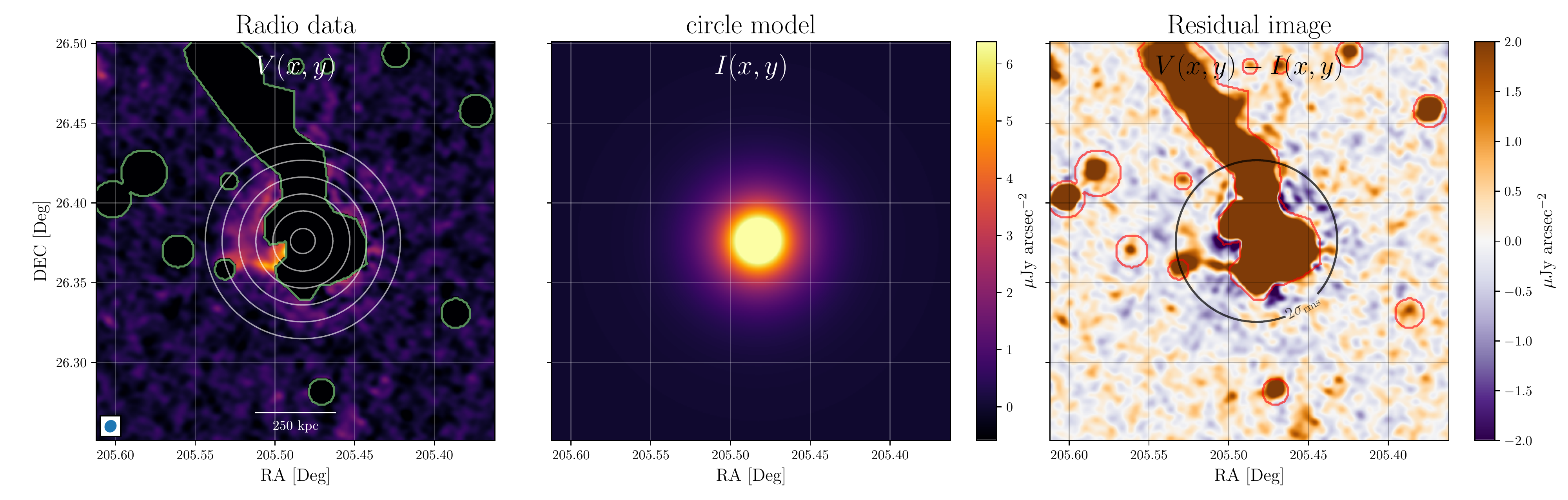} 
  \caption{Fitting result obtained with \textsc{Halo-FDCA} \citet{boxelaar21}. \textit{Left}: \lofar\ low-resolution image used for the fit. Contaminating regions are masked out and contoured in green. The circular contours show the model at $[1,2,4,\dots,32]\times\sigma$ levels. \textit{Center}: Image of the best-fit circular model. \textit{Right}: Residual image of the fit. The circle shows the $2\sigma$ level of the model. The red contours shows the masked regions. The $\chi^2$/d.o.f. of the fit is 1422.35/828. The \mcmc\ corner plot of the fit is reported in Appendix~\ref{app:error_map}.} 
 \label{fig:halo_profile}
\end{figure*}

Bright, irregular, diffuse radio emission is observed to the E of the two dominant elliptical galaxies of A1775 (Fig.~\ref{fig:optical_lofar}). Two main high-surface-brightness filaments can be identified in the \lofar\ image: F1 (accounting for $593\pm119$ mJy) is close to the brightest region of the head-tail radio galaxy, while F2 (accounting for $505\pm101$ mJy) is located in the southern region and elongated in the E-W direction. Diffuse emission is observed between these two structures, and this has a total flux density of $\simeq150$ mJy (this value is more uncertain due to residual artifacts in the image). This complex emission covers an area of $\sim230^2$ kpc$^2$ in projection, is detected in the \lofar\ 144~MHz and \gmrt\ 235~MHz images (but is well recovered only by the former), and not at higher frequencies (Fig.~\ref{fig:radio_images}). This suggests that it has a steep spectrum. Indeed, the distribution of the pixel values in the $144-235$~MHz spectral index map (Fig.~\ref{fig:spix}) for the two regions that cover the two main filaments have a consistent median of $\overline{\alpha} = 2.4$ (with standard deviation $\sigma_{\alpha} = 0.4$). This is in agreement with the non-detections at 610~MHz and 1.4~GHz and generally indicates that the emission has an ultra-steep spectrum ($\alpha>2$), whose precise value can only be determined with future deeper observation at $\gtrsim300$~MHz. Given the morphology of the radio source, its very steep spectrum, the lack of clear optical counterpart(s), and the possible co-location with some compression regions in the ICM (see end of Section~\ref{sec:connection}), we conclude that we are observing revived fossil plasma emission. An obvious candidate for the origin of these filaments is the double giant elliptical system in the core, that in the past may have injected the relativistic plasma in the ICM that has been subsequently revived by the gas dynamics triggered by motions in the core. \\
\indent
When the longest baselines of the \lofar\ observation are tapered down, further low-surface-brightness diffuse radio emission with roundish morphology is detected in the cluster center (Fig.~\ref{fig:chandra_lofar}). Although this newly discovered emission is strongly contaminated by the much brighter head-tail radio galaxy and the revived fossil plasma source, it is not obviously connected to any of them. This fainter emission is extended towards the NE, where it is remarkably confined by the arc-shaped cold front detected with \chandra\ (Section~\ref{sec:arc-shaped}), reported with the dashed yellow arc in Fig.~\ref{fig:chandra_lofar}.  \\
\indent
The flux density of the central diffuse radio emission, excluding the contamination of sources, is in the range of $70-80$ m\jy\ (depending on the adopted extracting region). It is likely that this diffuse emission also extends up to the regions where the emission of the head-tail radio galaxy and revived fossil plasma source is dominating. One could attempt to estimate the total flux density assuming that the central diffuse emission follows an exponential profile \citep[\eg,][]{murgia09} in the following form:

\begin{equation}\label{eq:exponential}
 I(r) = I_0 e^{-r/r_e}\:,
\end{equation}

\noindent
where $I_0$ is the central surface brightness and $r_e$ denotes the $e$-folding radius, and employing the recently developed Halo-Flux Density CAlculator\footnote{\url{https://github.com/JortBox/Halo-FDCA}} \citet[\textsc{Halo-FDCA}][]{boxelaar21}. This code allows us to estimate the flux density of extended diffuse sources such as giant and mini-halos in galaxy clusters. It performs a fit of the surface-brightness profile of the emission in 2D, assuming exponential profiles, allowing for regions of the image affected by contaminating sources to be masked and extrapolating the flux density beneath them. The fitting follows a \mcmcE\ (\mcmc) method to explore the parameter space. After masking the contribution of the sources embedded in the central diffuse emission of A1775, we performed the fit with \textsc{Halo-FDCA} assuming the circular model of Eq.~\ref{eq:exponential}. The low-resolution image used for the fit with the adopted masks, together with the best-fit model and residual image, are shown in Fig.~\ref{fig:halo_profile}. The best-fit values are $I_0=16.2\pm1.1$ $\mu$Jy arcsec$^{-2}$ and $r_e = 75\pm2$ kpc. The total flux density obtained integrating the profile up to $3r_e$ is $S_{144} = 245\pm50$ mJy. We stress that the majority of the diffuse flux is masked out, and the derived flux density relies on the validity of the extrapolation of the model in the masked regions, which cannot be tested. The measured flux density leads to a radio power of $P_{144} = 4 \pi D_L^2 S_{144}  = (3.1 \pm 0.7) \times 10^{24}$ \whz\ (we neglected the $k$-correction term due to the low-$z$ of the cluster and the unknown spectral index). The properties observed (\ie,\ source size, confinement of the emission in the cold front, and radio power) may indicate that the central diffuse emission is a radio mini-halo. However, as we see in Section~\ref{sec:dyn-state}, the classic definition of a mini-halo is challenged by the possible dynamical state of the cluster, and therefore in Section~\ref{sec:connection} we propose an alternative classification of the radio emission.

\section{Discussion}

\subsection{Dynamical state of the cluster and origin of the arc-shaped feature}\label{sec:dyn-state}

Literature studies do not agree on the dynamical state of A1775 \citep{andersson09, andradesantos17, lopes18, lagana19maps}. Although the cluster hosts a core that is colder than the surrounding medium (Fig.~\ref{fig:icm_maps}), it cannot be unambiguously defined as a cool-core or relaxed system. \citet{andradesantos17} used four metrics to identify cool-core clusters in a large sample of galaxy clusters observed with \chandra, and found that the values of A1775 (Tab.~\ref{tab:felipe}) are intermediate between cool-core and non-cool-core systems. Additionally, taking 3 keV as central temperature (Fig.~\ref{fig:icm_maps}) and using the central density reported in \citet{andradesantos17}, we estimate a central entropy $K_0 = kT n_{\rm core}^{-2/3} \simeq 52$ \kevcmsq, which is again close to the usual threshold adopted to separate the two classes of clusters \citep[\eg,][]{cavagnolo09}. The detailed view of the cluster central region provided by the new \chandra\ observation allowed us to highlight the presence of gas motions and X-ray discontinuities in the ICM (Section~\ref{sec:arc-shaped}). In view of our analysis, we offer two possible scenarios to explain the observed properties of A1775, hence its dynamical state.

\begin{table}
 \centering
 \caption{Concentration parameters ($C_{\rm SB}$ and $C_{\rm SB4}$), cuspiness ($\delta_{\rm cusp}$), and central gas density ($n_{\rm core}$) derived by \citet{andradesantos17} for A1775. Cool-core clusters typically have $C_{\rm SB} > 0.4$, $C_{\rm SB4} > 0.075$, $\delta_{\rm cusp} > 0.5$, and $n_{\rm core} > 0.015$ cm$^{-3}$.}
 \label{tab:felipe}
  \begin{tabular}{lcc} 
  \hline
  \hline
  Parameter & Value & Definition used \\
  \hline
  $C_{\rm SB}$ & $0.279  \pm 0.005$ & \citet{maughan12} \\%| break 0.4
  $C_{\rm SB4}$ & $0.0823 \pm 0.0021$ & \citet{santos08} \\%| break 0.075  $\delta_{\rm cusp}$ & $0.880 \pm 0.046$ & \citet{vikhlinin07} \\%| break 0.5
  $n_{\rm core}$ [cm$^{-3}$] & $0.01393 \pm 0.00065$ & -- \\%| break 0.015
  \hline
  \end{tabular}
\end{table}

In a "sloshing scenario", when a galaxy cluster is perturbed, its cool core starts sloshing in the cluster's potential well, generating a spiral of low-entropy, low-temperature gas, which is separated by the hot ICM by contact discontinuities \citep[\eg,][]{ascasibar06}. In A1775, we observe a spiral-pattern in the X-ray residual image of the best-fit surface-brightness profile, in the GGM-filtered image, and in the temperature and pseudo-entropy maps. The central diffuse radio emission confined by the arc-shaped cold front (Fig.~\ref{fig:chandra_lofar}) could then be interpreted as a radio mini-halo bounded by a sloshing cold front. We note that A1775 is part of a supercluster, indicating that it is in a dynamically active environment. This scenario could explain some of the properties of A1775 (\ie,\ the spiral of low-entropy gas and the arc-shaped cold front) and has also been advanced by \citet{lagana19maps}. Nonetheless, it cannot explain all the aspects of the system. In the classical sloshing scenario proposed by \citet{ascasibar06}, the cluster core should remain largely intact (\ie,\ compact and peaked in the X-rays), while it oscillates in the potential well. The core at the center of A1775 seems to be in a disruption phase and has a leading surface-brightness edge on the W (Fig.~\ref{fig:chandra_image}) that is reminiscent of a ``remnant-core'' cold front, rather than a sloshing cold front (see \citealt{tittely05} and \citealt{markevitch07rev, zuhone16rev} for reviews). Furthermore, the presence of two elliptical galaxies with similar luminosity and high radial-velocity difference \citep{chincarini71, jenner74} is difficult to reconcile with the scenario of a relaxed cluster with a sloshing core. \\
\indent 
In a "slingshot scenario", low-entropy or low-temperature gas spirals and arc-shaped features in the ICM can trace slingshot tails produced by off-axis mergers \citep[\eg, ][]{sheardown19}. The presence of a semi-disrupted core at the center of A1775 together with the two elliptical galaxies with a radial-velocity difference of $\sim 1800$ \kms\ \citep{chincarini71, jenner74} possibly suggest that A1775 underwent a more violent merger compared to that required for the sloshing scenario. In this case, the core that we are observing at the center of A1775 is that of a subcluster that fell into the system along the line of sight and with a nonzero impact parameter. The subcluster is now close to the apocenter of its orbit, producing a slingshot tail, and it is turning around for the second infall generating the leading edge on the W. The fact that there is only one core in A1775 can be explained if the main system did not host a cool core before the collision with the subcluster. Overall, the proposed scenario explains the features observed but requires a particular merger configuration and viewing angle. 

\begin{figure*}[h]
 \centering
 \includegraphics[width=\hsize,trim={0cm 0cm 0cm 0cm},clip,valign=c]{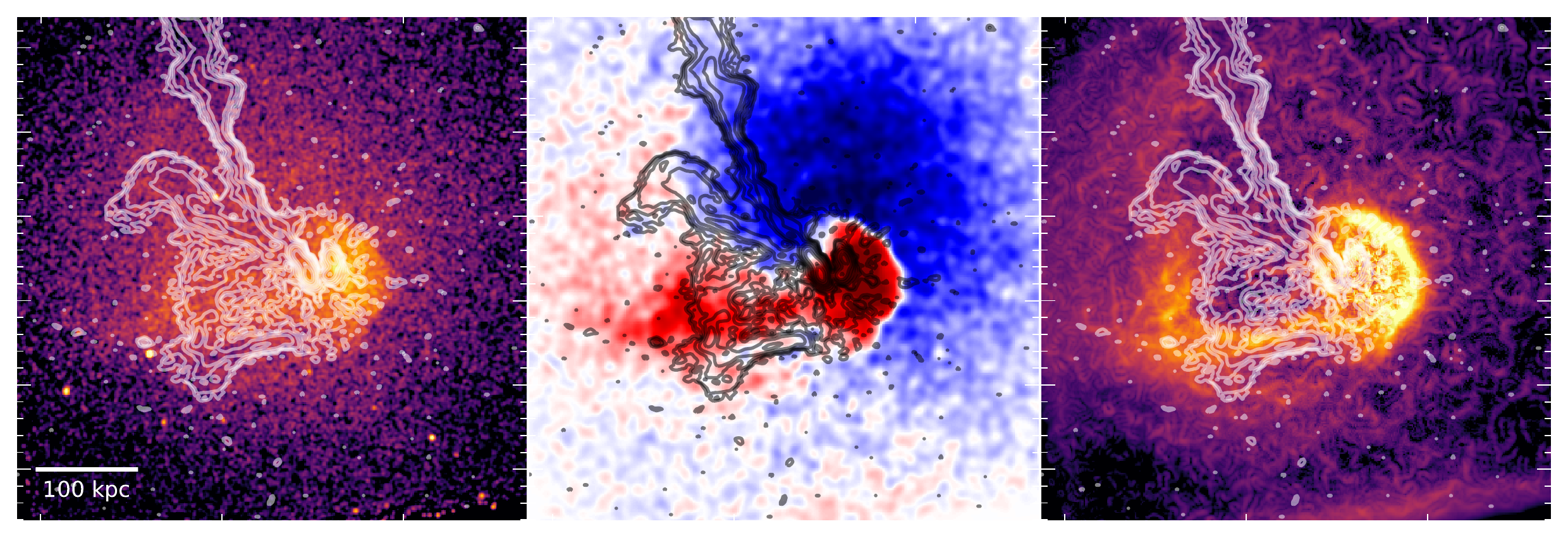}
  \caption{Zoom-in on the central cluster region. From left to right, the \lofar\ high-resolution contours of Fig.~\ref{fig:optical_lofar} are overlaid on the \chandra\ X-ray image, residual map, and GGM-filtered image of Figs.~\ref{fig:chandra_image} and \ref{fig:spiral}.}
 \label{fig:f1_f2}
\end{figure*}

Spiral-like structures and arc-shaped cold fronts in the ICM are typically generated by off-axis mergers. The collision with nonzero impact parameter is one of the conditions that generates sloshing motions and slingshot tails, which produce similar features and sometimes can be confused \citep[see Section 7 in][]{sheardown19}. At the moment, for A1775 we tend to prefer the slingshot scenario because: (i) the cluster core is not compact as expected in the case of ``classic'' sloshing\footnote{Here, we refer to the scenario of \citet{ascasibar06}, where a necessary condition for the onset of the sloshing is the presence of a steep entropy profile. We note, however, that sloshing has also been proposed in the literature for a few clusters without a cool core and with indications of merging activity, such as Abell 2142 \citep{rossetti13}, Abell 3560 \citep{venturi13a3560}, and Abell 1763 \citep{douglass18}.}, (ii) there is an X-ray discontinuity in the W that has the characteristics of a remnant-core cold front, and (iii) the cluster center has two elliptical galaxies with a strong velocity component along the line of sight. \\
\indent
X-ray spectral analysis can also provide information on the bulk motions of the ICM and therefore further constraints on the dynamical state of the system. However, the limited spectral resolution of CCDs makes this measurement really challenging and plagued by systematics. In the past, some detections, generally associated with large uncertainties, were claimed with \chandra\ \citep[\eg, ][]{dupke01centaurus, dupke06, dupke07a576}. More recently, \citet{liu15, liu16} proposed a strategy to infer bulk motions in clusters mainly with \acisi\ data once again recognizing, however, the difficulty of taking these measurements. In particular, the characterization of the time and spatial gain variation, such as that done with \xmm\ in the recent work by \citet{sanders20}, is crucial to performing the search for bulk motions. This type of analysis is very difficult to perform with \aciss\ and is beyond the scope of this paper. With all these caveats in mind, we attempted to measure the redshift from X-ray spectral analysis in different ICM regions but we were able to obtain reliable results only for the head of the mushroom (in a circular region with $r=40\arcsec$), where the fit provided $z=0.0719^{+0.003}_{-0.007}$. This value lies in the middle of the redshifts of the two elliptical galaxies ($z_{\rm B1339+266A} = 0.0757\pm 0.0001$ and $z_{\rm B1339+266B} = 0.0693 \pm 0.0001$, \citealt{davoust95}) and is possibly in line with the scenario of a merger along the line of sight. Outside the mushroom, the redshifts derived from the X-rays are basically unconstrained. In the future, the high-spectral resolution provided by microcalorimeters on-board upcoming X-ray missions such as \textit{XRISM} \citep{xrism20} and \athena\ X-IFU \citep{nandra13arx} will enable this kind of science. \\
\indent
The detailed analysis of the dynamical state of the cluster, the identification of optical/X-ray substructures, and the analysis of the main cold front in the W will be the subject of subsequent papers.

\begin{figure*}
 \centering
 \includegraphics[width=.62\hsize,trim={0cm 0cm 0cm 0cm},clip,valign=c]{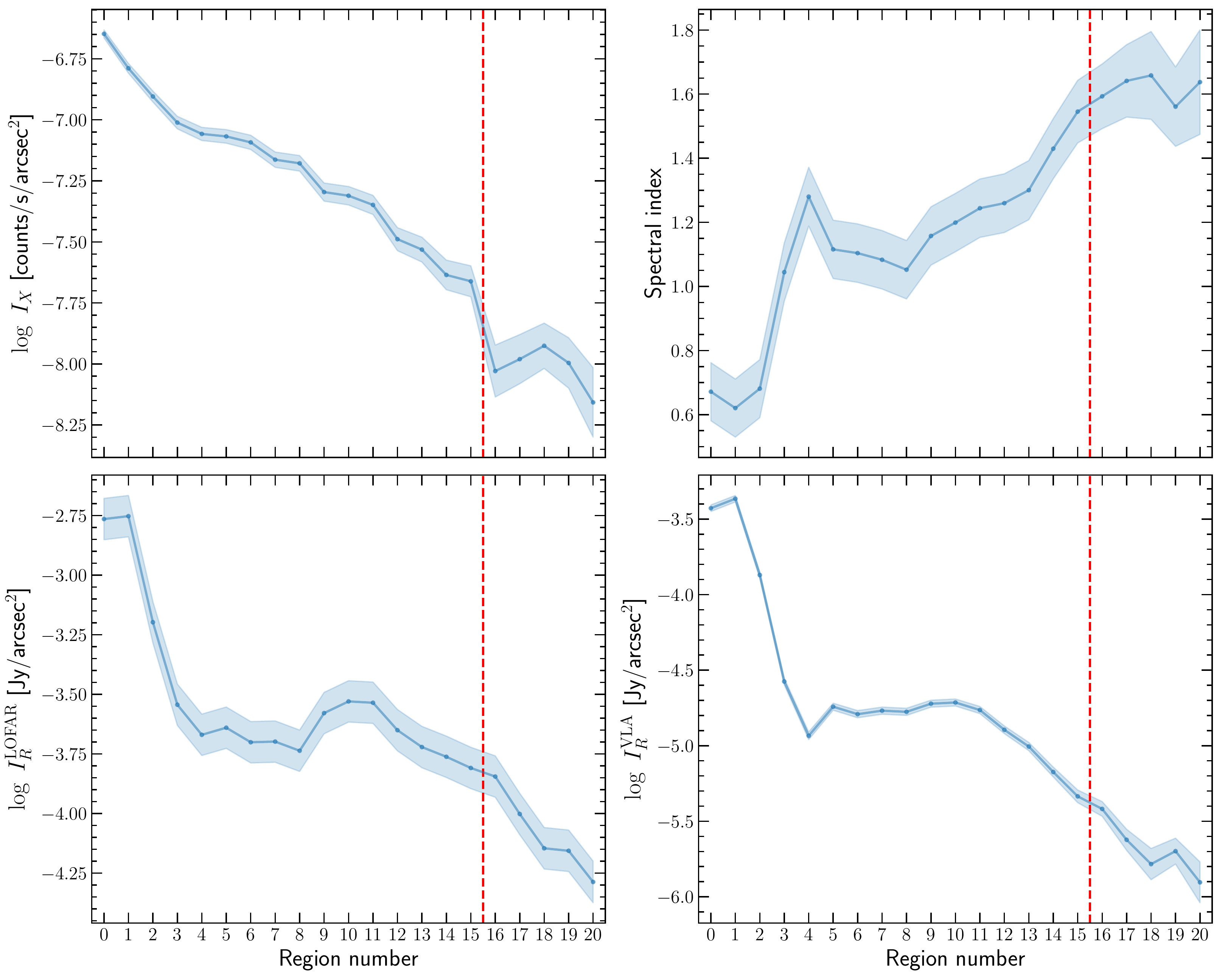}
 \includegraphics[width=.37\hsize,trim={3.55cm 0cm 3.55cm 0cm},clip,valign=c]{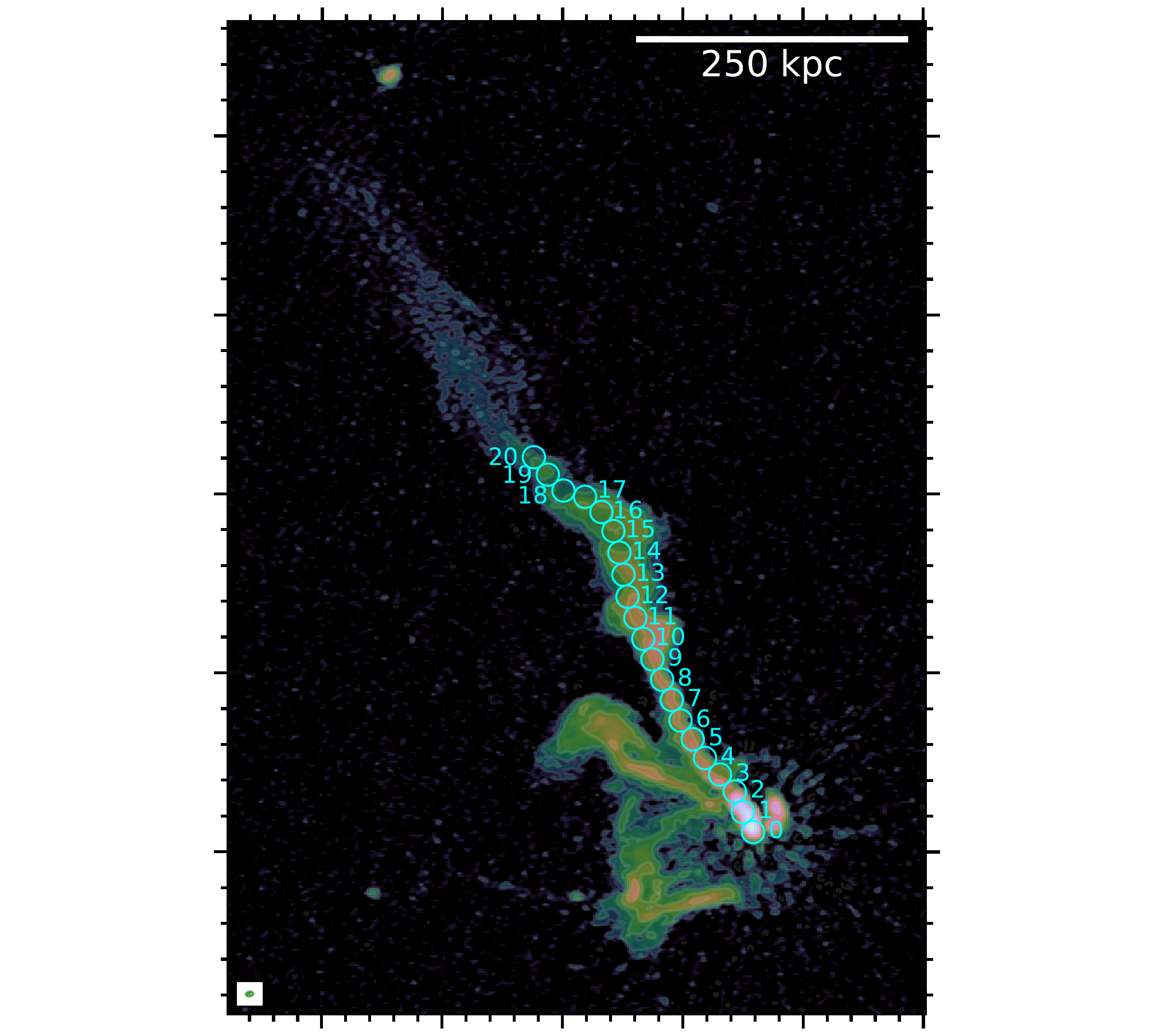}
  \caption{Spectral index, \lofar,\ and \vla\ radio surface brightness ($I_R$), and \chandra\ X-ray surface-brightness ($I_X$) trends along the head-tail radio galaxy. Radio surface brightnesses and spectral indexes were computed from \lofar\ and \vla\ C-array images with a beam of $15\arcsec\times15\arcsec$. Measurements were performed in the cyan beam-sized circular regions reported in the right panel, which depicts the same \lofar\ high-resolution image of Fig.~\ref{fig:optical_lofar} for visualization purposes only (the beam is shown in the bottom left corner). Regions are numbered from 0 (head of the tail) to 20 (last region where the \vla\ emission is $>3\sigma$). The red dashed vertical line on the left panels denotes the location of the arc-shaped cold front.}
 \label{fig:ptp}
\end{figure*}

\subsection{Connection between gas motions and central diffuse radio emission: a slingshot radio halo?}\label{sec:connection}

The determination of the dynamical state of a cluster is important to understand the origin of the extended synchrotron sources located in cluster centers, namely giant radio halos and radio mini-halos. While the former have a diffuse emission that generally extends over Mpc scales and are found in merging clusters, the latter cover the cooling region of relaxed systems and typically surround the AGN of the central cluster galaxy \citep[\eg, see][for a review]{vanweeren19rev}. Relativistic electrons in giant and mini-halos are believed to gain energy via merger-induced or sloshing-induced turbulent reacceleration processes, respectively \citep[\eg,][for a review]{brunetti14rev}. So, how does the central diffuse radio emission discovered in A1775 compare to this context? \\
\indent
On the one hand, the radio emission may resemble that of a mini-halo. First, it is located at the center of a cluster that has a distinct core with lower temperature or entropy than the ambient medium (Fig.~\ref{fig:icm_maps}). Second, it has an $e$-folding radius of $75\pm2$ kpc, while giant halos in the sample of \citet{murgia09} generally have $r_e > 100$ kpc (at 1.4~GHz). Third, it is confined by an arc-shaped cold front in the NE (Fig.~\ref{fig:chandra_lofar}), as is often observed in sloshing and mini-halo systems. On the other hand, mini-halos have only been found in systems with $K_0 < 30$ \kevcmsq\ \citep{giacintucci17} so far, while the central entropy of A1775 is $K_0 \simeq 52$ \kevcmsq. Moreover, in Section~\ref{sec:dyn-state} we argue that central gas sloshing, usually invoked to explain the origin of radio mini-halos, is not the preferred dynamical scenario for A1775. The small linear extent and relative high radio power, when compared to the low-mass of the system, point against the classification as giant radio halo (however, we note that to date giant radio halos have mainly been studied in massive systems undergoing major mergers, and their properties in low-mass clusters are still largely unconstrained). All this suggests that the extended emission should be classified in another way. \\
\indent
Pursuing the discussion on the cluster dynamical state (Section~\ref{sec:dyn-state}), we introduce and adopt the term ``slingshot radio halo'' for the central diffuse emission in A1775. The synchrotron emission appears mini-halo-like because sloshing and the slingshot develop under the same condition and produce similar features. It follows that sloshing spirals and slingshot tails are governed by the same physics. The difference between a mini-halo and a slingshot radio halo would then only be the dynamics that induced the gas motions in the ICM. \\
\indent
According to the turbulent reacceleration models, synchrotron-emitting electrons in mini-halos originate from a pre-existing population of seed relativistic electrons that are injected in the ICM by cluster AGN and that are spread and may be reaccelerated over $\lesssim500$ kpc by sloshing-induced turbulent motions \citep[\eg,][]{zuhone13}. The direct observations of these motions is very challenging, and, at the moment, constraints on the gas velocity dispersion are available only for the Perseus cluster, indicating that its ICM is sloshing at the level of a few hundred km s$^{-1}$ \citep{hitomi16, sanders20}. In slingshot tails, the atmosphere of the infalling system can be efficiently stripped near pericenter passage and become turbulent \citep[\eg,][]{roediger15b}, leading to the reacceleration of relativistic electrons via collisionless damping of turbulence. Arc-shaped sloshing or slingshot cold fronts are instead easier to observe and their origin is well understood. Beneath these discontinuities, shear magnetic amplification can increase the field strength \citep{keshet10distribution, zuhone11, reiss14}, leading to the stabilization of the fronts against Kelvin-Helmholtz instabilities. In turn, relativistic plasma can be restrained by this magnetic field and form diffuse radio emission bounded by X-ray edges. In A1775, the confinement of the slingshot radio halo within the discontinuity is remarkable (Fig.~\ref{fig:chandra_lofar}), and it is well in line with the scenario described above. \\
\indent
Seed relativistic electrons are a key ingredient of turbulent re-acceleration models as the direct acceleration of thermal pool electrons is a very inefficient mechanism. Cluster radio galaxies are a natural source of particles in the ICM. Interestingly, two radio loud elliptical galaxies are found at the center of A1775. One of the two is a giant head-tail radio galaxy that spans a projected size of about 800 kpc, meaning that it has the potential to inject relativistic electrons on a very large scale. In addition, we reported the presence of revived fossil plasma embedded in the diffuse radio emission which is direct evidence for the existence of seed particles in the environment. Therefore, the requirement of a pre-existing population of cosmic-ray electrons is clearly satisfied in this cluster. \\
\indent
The images shown in Fig.~\ref{fig:f1_f2} show a tantalizing connection between the radio emissions F1 and F2 and some features observed in the X-rays. In particular, F1 follows the ridge between the overdense and underdense region in the \chandra\ residual image (central panel) while F2 is coincident with a region where a strong X-ray surface-brightness gradient is present (right panel). It is possible that these old radio (light) filamentary structures passively trace the velocity field of the turbulent medium that is established by the interplay and mixing of the cold gas of the infalling subcluster with that of the main system. \\ 
\indent
The distortions observed on a linear scale $\sim 70$ kpc in F1 and F2 may result from shear motions in the velocity field that eventually generate a turbulent cascade in the ICM. These distortions suggest that the medium in the slingshot tail is quite turbulent with a velocity of the large-scale eddies that may be estimated as $\delta V_{\rm 1D} \sim L / t_{\rm cross}$ where $L \sim 70$ kpc and $t_{\rm cross}$ is the crossing time of the cold gas (CG), $t_{\rm cross} \sim D/V_{\rm CG}$, $D \sim 200$ kpc being the (projected) distance between the front of the leading edge (the cap of the mushroom) and the region where the distortions of F1 and F2 are developed. %
The resulting turbulent Mach number on about 70 kpc scale assuming isotropic turbulence and sound speed $c_{\rm s} \sim 1000$ \kms\ is $\mach \sim 0.3 \left(\frac{V_{\rm CG}}{500\:{\rm km/s}}\right)$. These conditions open to the possibility of turbulent acceleration with acceleration times of a few 100 Myr, which are not very different from those assumed in giant radio halos and radio bridges \citep[\eg,][]{brunetti16stochastic, brunetti20} and may explain the existence of steep spectrum diffuse radio emission on a scale of several hundreds of kpc.

\subsection{Extended head-tail}\label{sec:ht-analysis}

Our low-frequency observations with \lofar\ and \gmrt\ indicate that the head-tail radio galaxy is twice the size of the one previously reported (Fig.~\ref{fig:radio_images}), leading to a projected largest linear size of $\sim  800$ kpc. Interestingly, the newly discovered $\sim 400$ kpc extension discovered has different properties compared to the first part of the tail. It originates after a ``break'' of its structure, becoming more diffuse and wider, and shows an approximately constant surface brightness at 144~MHz (\ie,\ close to the break frequency suggested by Fig.~\ref{fig:radio_images}, assuming that adiabatic losses and strong variations of magnetic field do not play an important role), possibly indicating that the oldest population of electrons has been disturbed and reenergized, similarly to the case of the head-tail radio galaxies in Abell 1132 \citep{wilber18a1132} and ZwCl 0634.1+4750 (\citealt{cuciti18}, Cuciti et al., in prep.). The number of tailed radio galaxies showing this kind of feature is increasing thanks to the advent of highly sensitive observations at low frequencies, which are also possibly unveiling new gentle mechanisms of particle reenergization in tailed sources, as has been proposed for the case of the wide-angle radio galaxy in Abell 1033 \citep{degasperin17gentle}. As shown in Fig.~\ref{fig:chandra_lofar}, the tail breaks roughly at the location of the cold front detected in the X-rays. This suggests an interaction between the tail and the surrounding medium. Although the current data do not allow us to perform an accurate and detailed analysis of the spectral properties of the tail, the integrated spectra hint at the presence of a steep and curved spectrum for the newly discovered extension of the tail. \\
\indent
The analysis presented in Section~\ref{sec:results-ht} highlights the presence of a gradient along the tail, from $\alpha\sim0.6-0.7$ up to $\alpha\gtrsim1.5$, which is in agreement with that previously reported by \citet{giacintucci07}. In order to study the trends observed along the source structure in more detail, we produced the plots reported in Fig.~\ref{fig:ptp}. There, we compare the \chandra, \lofar, and \vla\ surface brightnesses, and the \lofar-\vla\ spectral index between 144~MHz and 1.4~GHz measured in the same 21 circular regions depicted in the right-hand panel following the ridge line of the head-tail where the emission is $>3\sigma$ in all observations. In order to increase the number of (beam-sized) regions for the analysis, we made use of a higher resolution \vla\ C-array image, and the final analysis was done on images convolved to the same resolution of $15\arcsec\times15\arcsec$. Below, we briefly comment each panel of Fig.~\ref{fig:ptp}. \\
\indent
The \chandra\ surface brightness ($I_X$) declines toward the end of the tail. This behavior is expected because the head of the tail is located in the core of the cluster, that is the X-ray brightest region of the ICM. Between regions 15 and 16, the surface brightness drops rapidly due to the presence of the cold front (see also Fig.~\ref{fig:NE_edge}). \\
\indent
In the \lofar\ surface-brightness profile ($I_R^{\rm LOFAR}$), the two innermost regions correspond to the head of the tail, where the radio surface brightness is highest. Then, the emission sharply decreases until it reaches region 4, and it remains roughly constant for five regions. Between regions 8 and 9, the tail gets brighter, and it fades again after region 11. Once the tail has crossed the X-ray discontinuity, its emission seems to decline faster. \\
\indent
The \vla\ surface brightness ($I_R^{\rm VLA}$) shows an initial trend similar to that observed in the \lofar\ image, with the difference that the surface brightness gets brighter between regions 4 and 5 before remaining roughly constant until region 11 (a small enhancement is noted between region 8 and 11, where the \lofar\ brightness shows a more pronounced ``bump''). Afterwards, the emission decreases regularly until it reaches the final region. \\
\indent
The spectral index has a fast steepening from $\alpha \sim 0.6-0.7$ in the inner regions up to $\alpha \sim 1.3$ in region 4. The spectral index then gets flatter ($\alpha \sim 1.1$) before steepening again reaching values up to $\alpha \sim 1.6$ in the last region. Within the uncertainties (shaded area), the steepening is consistently gradual along the tail. The investigation of possible variations in the spectral index trend (suggested by the mean values measured, such as between region 8 and 11) requires follow-up observations. \\
\indent
Once a relativistic jet has interacted with the ambient medium, it decelerates due to the entrainment of the external thermal gas \citep[\eg,][]{bicknell84, laing02dynamical, laing02relativistic} and flares due to the nonlinear growth of Kelvin-Helmholtz instabilities \citep[\eg,][]{jones79, loken95}. The position of the flaring point depends on how efficiently the interstellar medium of the host galaxy is stripped due to ram pressure during its transonic motion in the ICM \citep{gunn72}. In the samples of tailed sources presented by \citet{odonoghue93} and \citet{sebastian17}, the location of the flaring point varies from a minimum of $\sim15$ kpc, reaching values up to $\sim100$ kpc. These values may be affected by projection effects. For the head-tail in A1775, we found that the radio surface brightness shows two possible flaring points: between regions 4 and 5 at 1.4~GHz (where also a jump in spectral index index is present), and between regions 8 and 9 at 144~MHz (Fig.~\ref{fig:ptp}), corresponding to projected distances from region 0 of $\sim103$ kpc and $\sim185$ kpc, respectively. At such large distances, radio flaring can also be induced by dynamical motions in the ICM. The interplay between the head-tail radio galaxy and the thermal gas is evident where the structure of the tail breaks and changes direction. This transition occurs at the position of the arc-shaped cold font (\cf\ Fig.~\ref{fig:chandra_lofar}), where both a tangential gas velocity variation and a density gradient are present. The disruption and bending of the tail is expected in this case, where the development of Kelvin-Helmholtz instabilities can in principle reaccelerate the electrons and sustain their particle life-time \citep[\eg,][]{loken95}. Follow-up observations will shed light on the properties of the extension of the radio tail and test this scenario. We note that the possible interaction between a radio galaxy and a cold front has been recently claimed for NGC 4869 in the Coma cluster \citep{lal20ngc4869}. \\
\indent
Finally, we point out that in the first $\sim 400$ kpc the emission of the head-tail radio galaxy appears wobbly in the \lofar\ high-resolution image, similarly to the cases of IC 711 and NGC 4869 \citep[\eg,][for recent works]{sebastian17, wilber19, srivastava20, lal20ngc4869}. The widening and narrowing features indicate the development of instabilities along the tail, for example as a consequence of the turbulent wake generated behind the host galaxy while it moves at high speed in the ICM \citep[\eg,][]{jones79}.

\section{Conclusions}

We present a joint X-ray and radio analysis of the central region of A1775 ($z=0.072$) employing \chandra, \lofar\ 144~MHz, \gmrt\ 235~MHz and 610~MHz, and \vla\ 1.4~GHz observations. Our main findings are summarized as follows.

\begin{enumerate}
 \item We find clear evidence of gas motions in the cluster center and highlighted the presence of a cool, spiral-like structure in the thermal ICM. This pattern is observed in the X-ray surface brightness and in the temperature and pseudo-entropy maps.
 
 \item At the NE boundary of the gas spiral, the X-ray surface brightness drops. The broken power-law fit indicates that the gas density changes by a factor $\compr \simeq 1.7$ across the jump. The inner and outer temperatures are $kT_{\rm in} = 3.0^{+0.3}_{-0.4}$ keV and $kT_{\rm out} = 3.9^{+1.5}_{-0.9}$ keV respectively, suggesting that the edge is tracing a cold front.
 
 \item We proposed two scenarios to explain the origin of the arc-shaped feature revealed by \chandra. We argued that the preferred scenario is that of an off-axis merger along the line of sight, where the infalling subcluster has passed the pericenter and is turning around, producing a slingshot tail of low-entropy and low-temperature gas. Follow-up work is required to confirm the suggested dynamical configuration of the system.
 
 \item Owing to the low-frequency observations performed with \lofar\ and \gmrt, we detected a $\sim 400$ kpc extension of the head-tail radio galaxy in A1775, leading to a total (projected) extent of this source of $\sim 800$ kpc, doubling its previously reported size. The structure of the tail is collimated for the first $\sim 400$ kpc, then it breaks and becomes more diffuse. This transition occurs at the location of the arc-shaped cold front. We speculate that the outer region of the tail originates from the re-acceleration of dormant tail electrons; deeper observations will shed light on this point.
 
 \item At low frequencies, filamentary and diffuse radio emission is detected in the cluster center. The emission is recovered well only by \lofar\ and is present also in the \gmrt\ 235~MHz data, suggesting an ultra-steep spectrum of $\alpha > 2$. Due to its spectral properties, morphology, lack of clear optical counterpart(s) and possible connection with compression regions in the ICM, we classified it as revived fossil plasma.
 
 \item We discovered extended, diffuse emission in the cluster center. This claim is solely based on the \lofar\ low-resolution image. The source is bounded by the arc-shaped cold front. Under the assumption of a circular exponential model, the derived flux density at 144~MHz is $S_{144} = 245\pm50$ mJy, corresponding to a radio power of $P_{144} = (3.1 \pm 0.7) \times 10^{24}$ \whz. In line with the proposed dynamical scenario, we speculated that this central diffuse radio emission is tracing a slingshot radio halo. An intriguing indication of significant turbulent motions in the central region of the cluster is provided by the distortions observed in the filamentary structures F1 and F2.
\end{enumerate}

Overall, our work indicates the presence of substantial dynamical gas motions in the central region of A1775. The discovery of a broken head-tail, revived fossil plasma, and a diffuse central emission probes ongoing interplay between thermal and nonthermal components in the ICM and particle reacceleration mechanisms in the cluster center.

\paragraph{\textit{Note added in proof.}} After this paper was accepted, \citet{hu21arx} made available results based on the analysis of \chandra\ and \xmm\ observations of Abell 1775. Their findings are in good agreement with those presented in this paper. They also used numerical simulations and suggested that the X-ray properties observed in Abell 1775 could be produced by what we refer to as a sloshing scenario.

\begin{acknowledgements}
We thank the anonymous referee for the useful comments on the manuscript.AB and RJvW acknowledge support from the VIDI research programme with project number 639.042.729, which is financed by the Netherlands Organisation for Scientific Research (NWO).
FG, GB, RC, and MR acknowledge support from INAF mainstream project `Galaxy Clusters Science with LOFAR' 1.05.01.86.05.
VC acknowledges support from the Alexander von Humboldt Foundation.
AD acknowledges support by the BMBF Verbundforschung under the grant 05A17STA.
LOFAR \citep{vanhaarlem13} is the LOw Frequency ARray designed and constructed by ASTRON. It has observing, data processing, and data storage facilities in several countries, which are owned by various parties (each with their own funding sources), and are collectively operated by the ILT foundation under a joint scientific policy. The ILT resources have benefitted from the following recent major funding sources: CNRS-INSU, Observatoire de Paris and Universit\'{e} d'Orl\'{e}ans, France; BMBF, MIWF-NRW, MPG, Germany; Science Foundation Ireland (SFI), Department of Business, Enterprise and Innovation (DBEI), Ireland; NWO, The Netherlands; The Science and Technology Facilities Council, UK; Ministry of Science and Higher Education, Poland; Istituto Nazionale di Astrofisica (INAF), Italy. This research made use of the Dutch national e-infrastructure with support of the SURF Cooperative (e-infra 180169) and the LOFAR e-infra group, and of the LOFAR IT computing infrastructure supported and operated by INAF, and by the Physics Dept.~of Turin University (under the agreement with Consorzio Interuniversitario per la Fisica Spaziale) at the C3S Supercomputing Centre, Italy. The J\"{u}lich LOFAR Long Term Archive and the German LOFAR network are both coordinated and operated by the J\"{u}lich Supercomputing Centre (JSC), and computing resources on the supercomputer JUWELS at JSC were provided by the Gauss Centre for Supercomputing e.V. (grant CHTB00) through the John von Neumann Institute for Computing (NIC). This research made use of the University of Hertfordshire high-performance computing facility and the LOFAR-UK computing facility located at the University of Hertfordshire and supported by STFC [ST/P000096/1]. We thank the staff of the GMRT for making these observations possible. GMRT is run by the National Centre for Radio Astrophysics of the Tata Institute of Fundamental Research. The NRAO is a facility of the National Science Foundation operated under cooperative agreement by Associated Universities, Inc. SRON is supported financially by NWO, The Netherlands Organisation for Scientific Research. Basic research in radio astronomy at the Naval Research Laboratory is supported by 6.1 Base funding. The scientific results reported in this article are based in part on data obtained from the \chandra\ Data Archive. This research made use of APLpy, an open-source plotting package for Python \citep{robitaille12}.
\end{acknowledgements}

\bibliographystyle{aa}
\bibliography{library.bib}

\begin{appendix}

\section{Spherical vs. elliptical $\beta$-models}\label{app:elliptical_model}

\begin{figure}
 \centering
 \includegraphics[width=\hsize,trim={0cm 0cm 0cm 0cm},clip]{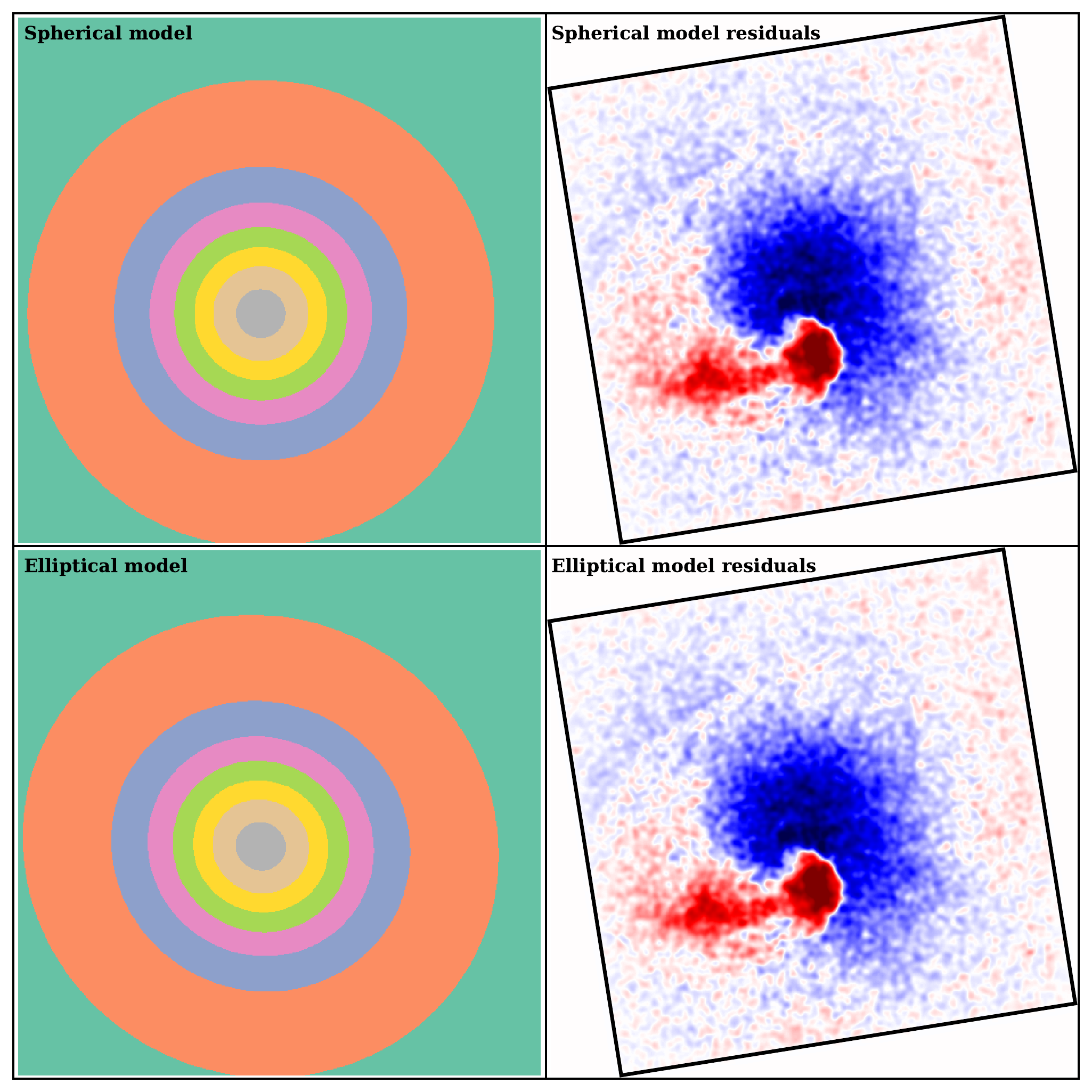}
  \caption{Visual comparison between best-fit spherical and elliptical $\beta$-models (left panels) and their residuals (right panels).}
 \label{fig:spherical_vs_elliptical}
\end{figure}

By using the \ciao\ task \texttt{dmellipse}, we found that an ellipse that encloses 90\% of the ICM photons has a ratio between major and minor axis of 1.05 and orientation angle of 333\deg. Due to the small axis ratio, an elliptical $\beta$-model with this shape does not provide any significant difference with respect to the spherical model adopted in the main text, and shows the same spiral-like pattern in the residual image (Fig.~\ref{fig:spherical_vs_elliptical}). The best-fit elliptical model has a core radius of $r_{\rm c} = 2.63\pm0.05$ arcmin and $\beta = 0.84\pm0.01$, which are consistent with the values reported in Section~\ref{sec:arc-shaped} for the spherical model.

\section{Error maps and corner plot}\label{app:error_map}

\begin{figure}
 \centering
 \includegraphics[width=\hsize,trim={1.5cm 0cm 1.5cm 0cm},clip]{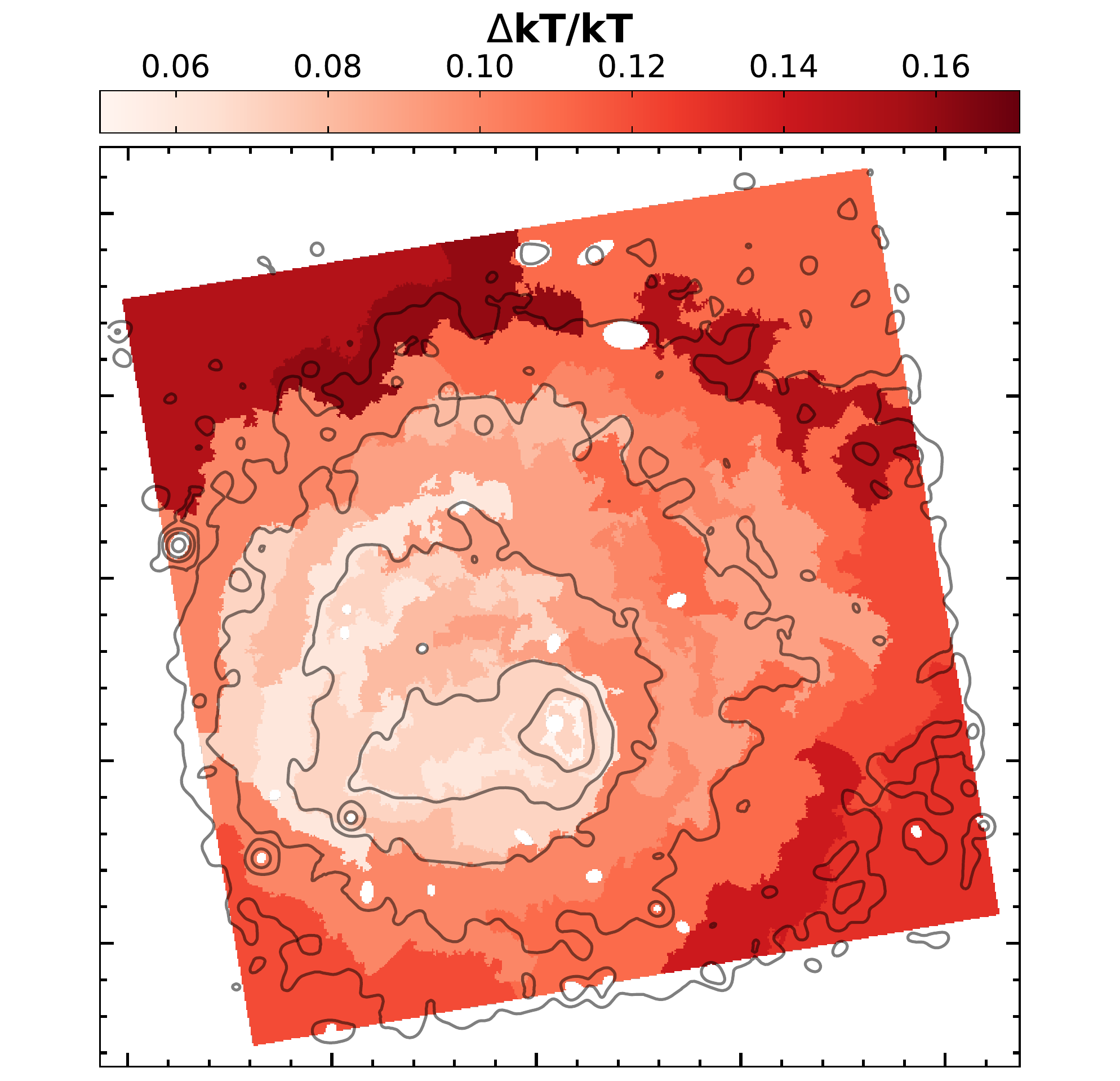}
  \caption{Temperature error map corresponding to Fig.~\ref{fig:icm_maps}.}
 \label{fig:kt_error}
\end{figure}

\begin{figure}
 \centering
 \includegraphics[width=\hsize,trim={0cm 0cm 0cm 0cm},clip]{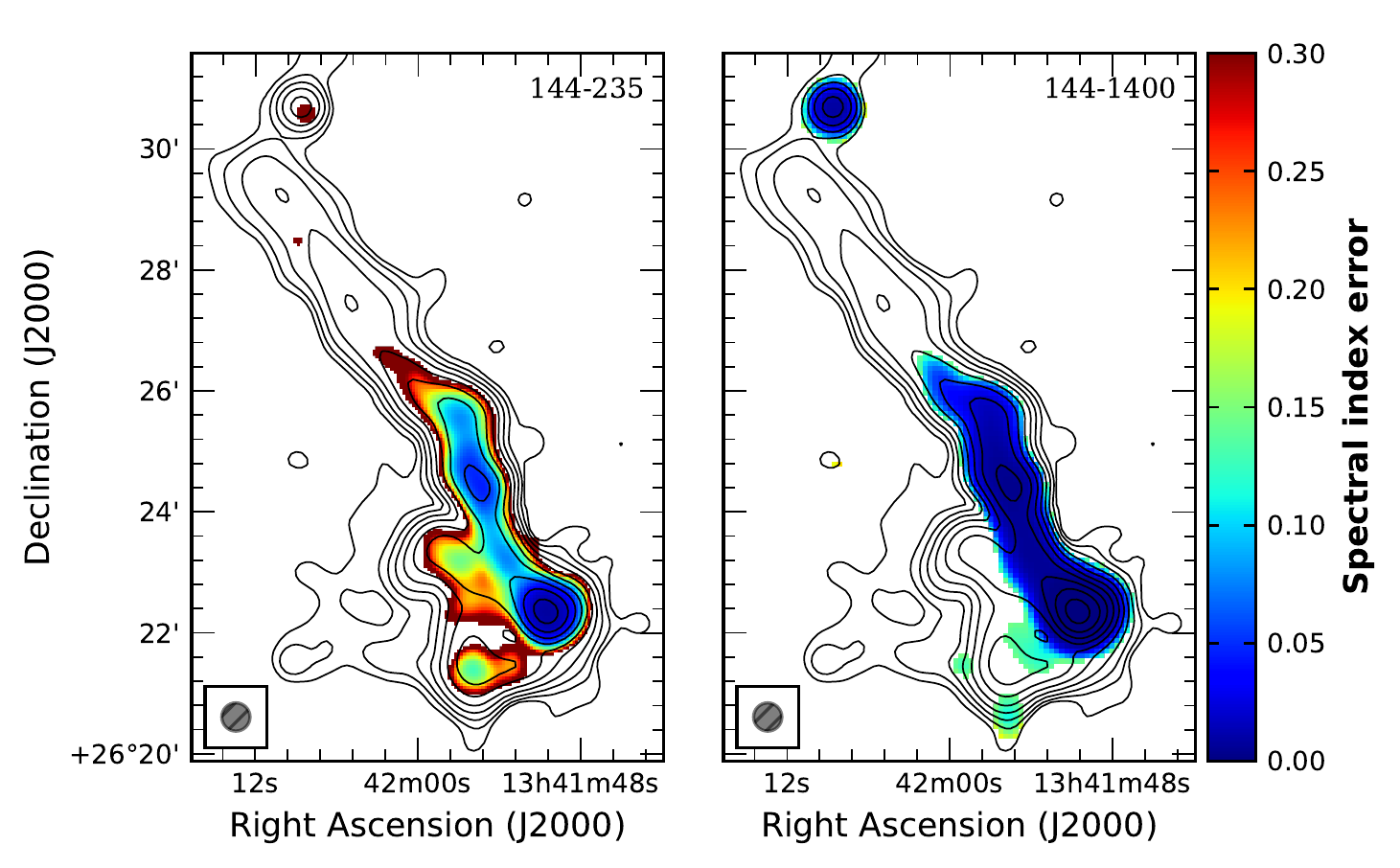} 
  \caption{Spectral index error maps corresponding to Fig.~\ref{fig:spix}.}
 \label{fig:spix_error}
\end{figure}

\begin{figure*}
 \centering
 \includegraphics[width=\hsize,trim={0cm 0cm 0cm 0cm},clip]{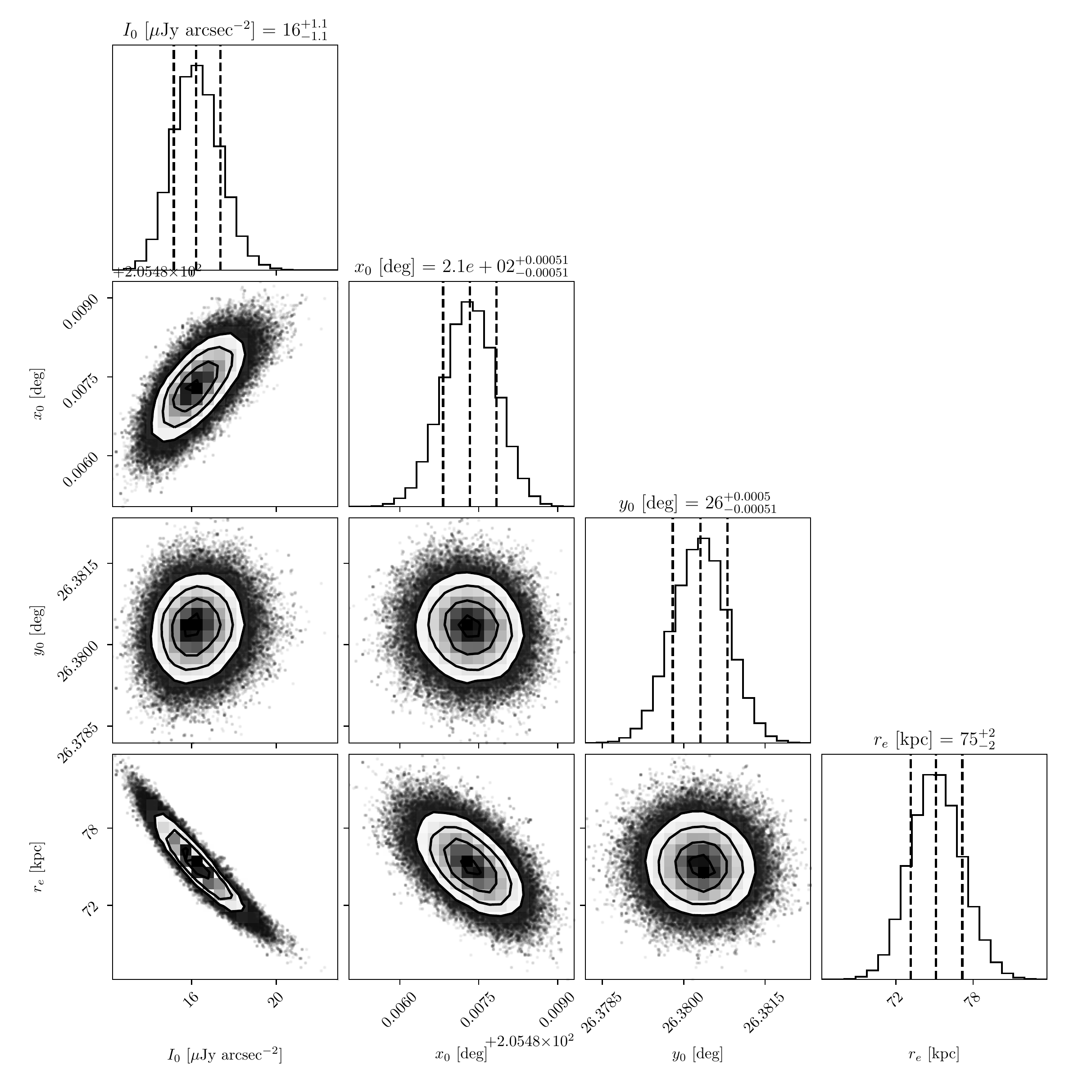} 
  \caption{\mcmc\ corner plot for the central diffuse emission surface-brightness profile fit corresponding to Fig.~\ref{fig:halo_profile}.}
 \label{fig:corner_plot}
\end{figure*}

The temperature error map is reported in Fig.~\ref{fig:kt_error}. The pseudo-pressure and pseudo-entropy maps have similar fractional errors given the small uncertainty (a few percent) associated to the emission measure from which they depend \citep[see][]{botteon18edges}. The low- and high-frequency spectral index error maps are reported in Fig.~\ref{fig:spix_error}. \\
\indent
The \mcmc\ corner plot \citep{foremanmackey13} showing the distribution of solutions for the fit parameters of the central diffuse emission surface-brightness profile is shown in Fig.~\ref{fig:corner_plot}.

\end{appendix}

\end{document}